\documentclass[aps,prd,nofootinbib,floatfix,showpacs,notitlepage,eqsecnum,10pt]{revtex4-2} 

\usepackage{graphicx,slashed,bbold,bm}
\usepackage{amsmath,amssymb,epsfig}
\usepackage[dvipsnames]{xcolor}
\usepackage{braket}
\usepackage{epstopdf}
\usepackage{multirow}
\usepackage{booktabs}
\usepackage{ragged2e}
\usepackage[colorlinks=true,linkcolor=blue,urlcolor=blue,citecolor=blue]{hyperref}
\usepackage{cleveref}
\usepackage{simpler-wick}
\usepackage{chngcntr}
\counterwithout{equation}{section}
\usepackage{makecell}
\usepackage{needspace}
\usepackage{tikz}
\newcommand*\circled[1]{\tikz[baseline=(char.base)]{
    \node[shape=circle,draw,inner sep=1pt] (char) {#1};}}

\newcommand{\bi}{\begin{itemize}}
\newcommand{\ei}{\end{itemize}}

\newcommand{\nc}{\newcommand}
\nc{\non}{\nonumber}
\nc{\hc}{\hbox {H.c.}}
\nc{\noi}{\noindent}
\nc{\barx}{\bar{x}}
\nc{\pbarn}{\;\hbox {pb}}
\nc{\fbarn}{\;\hbox {fb}}
\nc{\cone}{{\scriptsize \textcolor{orange}{\circled{1}}}}
\nc{\ctwo}{{\scriptsize \textcolor{orange}{\circled{2}}}}
\nc{\cthr}{{\scriptsize \textcolor{orange}{\circled{3}}}}

\nc{\hsp}{\hspace{0.5cm}}
\nc{\lsp}{\hspace{1cm}}
\nc{\Lsp}{\hspace{2cm}}
\nc{\LLsp}{\lsp\lsp}
\nc{\lra}{\longrightarrow}
\nc{\p}{\prime}
\nc{\sgn}{\text{sgn}}
\nc{\ph}{\varphi}
\nc{\eff}{\mathrm{eff}}
\nc{\sqM}{\sqrt{M}}
\nc{\NL}{\mathrm{NL}}
\nc{\moc}{\mathcal{M}} %Meson octet
\nc{\rd}[1]{\textcolor{red}{#1}}
\nc{\bl}[1]{\textcolor{blue}{#1}}
\nc{\pbar}{\bar{\psi}}

\nc{\vp}[1]{\mathbf{p}_{#1}}
\nc{\vk}{{\bm k}_\perp}
\nc{\vkp}{{\bm k}'_\perp}
\nc{\vq}{{\bm q}_\perp}
\nc{\uu}{{\uparrow\uparrow}}
\nc{\ud}{{\uparrow\downarrow}}
\nc{\du}{{\downarrow\uparrow}}
\nc{\dd}{{\downarrow\downarrow}}
\def\lsim{\mathrel{\raise.3ex\hbox{$<$\kern-.75em\lower1ex\hbox{$\sim$}}}}
\def\gsim{\mathrel{\raise.3ex\hbox{$>$\kern-.75em\lower1ex\hbox{$\sim$}}}}

\def\udots{\mathinner{\mkern1mu\raise1pt\vbox{\kern7pt\hbox{.}}\mkern2mu\raise4pt\hbox{.}\mkern2mu\raise7pt\hbox{.}\mkern1mu}}

\def\vp{{\bf p}}

\def\be{\begin{equation}}
\def\ee{\end{equation}}
\def\bea{\begin{eqnarray}}
\def\eea{\end{eqnarray}}
\def\vp{{\bf p}}

\def\be{\begin{equation}}
\def\ee{\end{equation}}
\def\bea{\begin{eqnarray}}
\def\eea{\end{eqnarray}}

\newcommand{\revCbar}[1]{{#1}}
\begin{document}
% Preprint placeholder removed in review.

\title{Gravitational form factors of heavy quarkonia and their
nonrelativistic expansion in a self-consistent light-front quark model}

\author{Ho-Meoyng Choi}
\email[ E-mail: ]{homyoung@knu.ac.kr}
\affiliation{Department of Physics,
 Teachers College, Kyungpook National University, Daegu 41566, Republic of Korea}

\author{Chueng-Ryong Ji}
\email[ E-mail: ]{crji@ncsu.edu}
\affiliation{Department of Physics and Astronomy, North Carolina State University, Raleigh, NC 27695-8202, USA}

\author{Hyeon-Dong Son}
\email[ E-mail: ]{hdson@inha.ac.kr}
\affiliation{Department of Physics,
 Inha University, Incheon 22212, Republic of Korea}
  
\begin{abstract}
We investigate the gravitational form factors of the pseudoscalar
heavy quarkonia $\eta_c(1S,2S)$ and $\eta_b(1S,2S)$ in a light-front
quark model with physical radial wave functions represented by
orthogonal mixtures of the two lowest harmonic-oscillator basis
states $\phi_{1S}$ and $\phi_{2S}$.
The Bakamjian--Thomas prescription is applied consistently to
constituent matrix elements and Lorentz prefactors at the integrand
level.
Working in the $\Delta^+=0$ frame with $\Delta=P'-P$ and
$t=\Delta^2$, we extract $A_q(t)$ from the quark energy--momentum
tensor components $T_q^{++}$, $T_q^{+i}$, and $T_q^{ij}$, and
$D_q(t)$ from $T_q^{+-}$, $T_q^{-i}$, and $T_q^{ii}$ using
appropriate momentum-transfer projections.
The independent extractions agree in the full relativistic
calculation and in the analytic nonrelativistic expansions through
$\mathcal O(v^2)$, where $v=\beta/m_Q$, $\beta$ is the Gaussian
width, and $m_Q$ is the constituent mass.
At zero momentum transfer, the expansions give
$D_q^{\phi_{1S}}(0)=-v^2+\mathcal O(v^4)$ and
$D_q^{\phi_{2S}}(0)=-2/3-(13/3)v^2+\mathcal O(v^4)$.
The interference, defined as the sum of the two cross overlaps,
first contributes as
$D_q^{\mathrm{cross}}(0)=-(4/\sqrt{6})v^2+\mathcal O(v^4)$.
Thus, the pure Gaussian $1S$ contribution is suppressed, whereas
a nonzero $2S$ admixture at fixed mixing angle leaves an
$\mathcal O(v^0)$ D-term in the physical ground state.
Full numerical results for the mixed states illustrate the effects
of heavy-quark flavor and radial structure on mass distributions
and internal stresses.
\end{abstract}

\maketitle

\section{Introduction}\label{Sec:I}

The hadronic matrix elements of the total energy-momentum tensor (EMT)
$T^{\mu\nu}=T_q^{\mu\nu}+T_g^{\mu\nu}$ of quantum chromodynamics (QCD)
encode information about the mass and momentum distributions within
hadrons~\cite{Kobzarev:1962wt,Pagels:1966zza}.
For a spin-0 hadron, the matrix elements of this conserved tensor are
parameterized by two gravitational form factors (GFFs), $A(t)$ and $D(t)$.
While momentum conservation and relativistic state normalization
fix $A(0)=1$, $D(0)$ is not constrained by any general conservation law.
Since the tensor structure multiplying $D(t)$ vanishes at zero momentum
transfer, off-forward matrix elements are needed to determine this form
factor and resolve the internal pressure and shear-force distributions
that it governs~\cite{Polyakov:2002yz}.
These stress distributions are commonly analyzed using the global
von~Laue condition and the local normal-force stability
conditions~\cite{Lorce:2018egm,Panteleeva:2020ejw}.
Comprehensive reviews are given in
Refs.~\cite{Polyakov:2018zvc,Lorce:2025oot}.

Since direct gravitational measurements are not feasible at hadronic scales,
experimental access to GFFs relies on indirect probes.
A principal connection is provided by the Mellin moments of twist-2
vector generalized parton distributions (GPDs)~\cite{Muller:1994ses}.
The first moments of quark GPDs give flavor vector form factors, whose
charge-weighted sum yields the electromagnetic form factor, while the
second moments of the GPDs give combinations of GFFs.
For a spin-0 target, the $D$-term contribution to the second moment is
proportional to the squared skewness $\xi^2$, so determining $D(t)$
through this moment requires $\xi\neq0$~\cite{Polyakov:2018zvc}.
% However, $D(t)$ can also be determined directly from suitable
% projections of off-forward EMT matrix elements, even at zero
% skewness with nonzero transverse momentum transfer. 

With the free-field EMT conventions of
Refs.~\cite{Hudson:2017xug,Hudson:2017oul}, a free Klein--Gordon
boson has $D_{\rm KG}(t)=-1$, whereas a free Dirac fermion has
$D_{\rm Dirac}(t)=0$.
The soft-pion theorem gives the same forward value,
$D_\pi(0)=-1$, for the total pion EMT in the chiral
limit~\cite{Polyakov:1999gs}, but this equality does not imply
identical off-forward form factors.
The estimated leading chiral correction reduces
$|D_\pi(0)|$~\cite{Donoghue:1991qv,Hudson:2017xug},
consistent with the monopole-fit lattice result
$D_\pi(0)=-0.900(70)$ at $m_\pi\approx170$~MeV~\cite{Hackett:2023nkr}.

Beyond the pion, the $D$-term has been studied across a wide range
of systems.
For the nucleon, calculations have been performed in model and
continuum approaches~\cite{Goeke:2007fp,Wakamatsu:2007uc,Cebulla:2007ei,
Kim:2012ts,Jung:2013bya,Neubelt:2019sou,Kim:2020nug,Yao:2024ixu}.
Phenomenological analyses access quark $D$-term information through
deeply virtual Compton scattering~\cite{Burkert:2018bqq}
and gluon $D$-term information through near-threshold heavy-quarkonium
production~\cite{Guo:2023pqw,Wang:2023fmx,Duran:2022xag},
subject to the assumptions of the respective extraction frameworks.
Complementary determinations come from dispersive
analyses~\cite{Pasquini:2014vua,Cao:2024zlf}
and lattice QCD~\cite{Hackett:2023rif}.
Calculations of the nuclei generally find $D(0)<0$, with $|D(0)|$
growing strongly overall with mass number~\cite{Polyakov:2002yz,
Guzey:2005ba,GarciaMartin-Caro:2023klo,GarciaMartin-Caro:2023toa},
although shell effects produce departures from monotonic
behavior~\cite{Hatta:2026upg}.
Microscopic studies also address the lightest
nuclei~\cite{He:2023ogg,He:2024vzz,He:2024jgc}.
Spin-1 targets such as the deuteron require several $D$-type
form factors rather than a single one~\cite{Cosyn:2019aio,
Polyakov:2019lbq,Freese:2022yur,Cosyn:2026gyy}.
Negative $D$-terms are also found in solitonic configurations
such as Q-balls and Q-clouds~\cite{Mai:2012yc,Mai:2012cx,Cantara:2015sna}.
Positive monopole $D$-terms have been reported for
spin-$3/2$ baryons in quark--diquark
models~\cite{Fu:2022rkn,Wang:2023bjp},
and positive values have also been obtained for the hydrogen
atom~\cite{Czarnecki:2023yqd,Ji:2022exr}.

For heavy quarkonia, existing calculations yield substantially different
$D$-terms.
For the pseudoscalar charmonium ground state $\eta_c(1S)$,
basis light-front quantization (BLFQ) gives
$D_{\eta_c}(0)\sim-5$ using the light-front energy density $T^{+-}$
and the shear stress $T^{12}$~\cite{Xu:2024hfx,Hu:2024edc},
whereas a contact-interaction calculation based on coupled
Dyson--Schwinger and Bethe--Salpeter
equations~\cite{Sultan:2024hep} yields $D_{\eta_c}(0)=-0.576(3)$.
Lattice QCD calculations of charmonia have so far been restricted
to forward matrix elements of the energy-momentum tensor, such as
the quark--gluon mass decomposition and momentum
fractions~\cite{Yang:2014xsa,Sun:2020pda,Bollweg:2026nnr}, 
so no lattice $D$-term is yet available for comparison.
These differing predictions motivate a closer examination of how
bound-state wave functions and the treatment of EMT matrix elements
enter the quarkonium $D$-term.

For nonrelativistic heavy quarkonium with
$m_Q\gg\Lambda_{\rm QCD}$, the hierarchy underlying nonrelativistic
QCD consists of the hard ($m_Q$), soft ($m_Qv$), and
ultrasoft ($m_Qv^2$) scales, where $v\ll1$ characterizes the
heavy-quark velocity in the bound state
($v_c^2\sim0.3$, $v_b^2\sim0.1$)~\cite{Brambilla:2004jw}.
This hierarchy motivates a velocity expansion of the GFFs to
identify contributions that survive in the nonrelativistic limit
and distinguish them from relativistic corrections.
Comparing the $1S$ and $2S$ states of charmonium and bottomonium
within a common framework further allows the roles of the heavy-quark
mass scale and radial structure to be examined.

The light-front (LF) approach provides a relativistic framework
for this analysis through overlap representations of hadronic matrix
elements in terms of LF wave functions (LFWFs), whose internal
variables are invariant under kinematic LF boosts~\cite{BD2001}.
This representation allows the full relativistic form factors and
their nonrelativistic expansions to be compared using the same
wave functions.
At zero skewness, the transverse and minus EMT components used to
extract the $D$-term can be sensitive to LF zero-mode contributions.
Agreement among extractions from different EMT components therefore
provides an important internal consistency test of the calculation.

The framework used here is a self-consistent light-front quark model
(LFQM) based on the Bakamjian--Thomas (BT)
construction~\cite{BT53,Poly10,CCKP}.
Its defining prescription replaces the physical meson masses by the
corresponding free-constituent invariant masses in both the constituent
matrix elements and the kinematic Lorentz structures at the integrand
level.
We refer to this uniform implementation as the full BT-based (fBT)
LFQM and describe it in Sec.~\ref{Sec:II}.
Earlier studies show that applying the replacement only to the matrix
elements can leave residual current-component dependence in
zero-mode-sensitive channels~\cite{Choi:2024ptc,Choi:2025kaon}.
The self-consistent LFQM framework has been applied to decay constants
and distribution amplitudes through higher twist~\cite{CJ14,CJ15,CJ17,
Jafar1,Jafar2}, semileptonic and rare transition form
factors~\cite{Choi21,ChoiAdv}, and electromagnetic and scalar form
factors and unpolarized $T$-even transverse-momentum-dependent
distributions of the pion and
kaon~\cite{Choi:2024ptc,Choi:2025kaon}.

Extending the pion GFF study~\cite{Choi:2025}, we consider
$\eta_c(1S,2S)$ and $\eta_b(1S,2S)$ using spectroscopy-constrained
orthogonal mixtures of the two lowest harmonic-oscillator (HO)
basis states.
We evaluate the constituent EMT in the valence $Q\bar Q$ sector,
with binding dynamics encoded in the mass operator rather than
an explicit gluon EMT.
Here $A_q(t)$ and $D_q(t)$ denote the contribution of one constituent
quark.
The antiquark contribution is identical for the equal-mass systems
considered here, giving the total valence form factors
$A(t)=2A_q(t)$ and $D(t)=2D_q(t)$.

{Working in the $\Delta^+=0$ frame, we compare three current-component extractions 
of each GFF in the full relativistic calculation. 
The form factor $\bar c_q(t)$ is extracted independently
and found to vanish within the fBT prescription.}

We derive complete analytic nonrelativistic (NR) expansions of both
GFFs through $\mathcal O(v^2)$, including the diagonal HO contributions
and their interference.
In this model expansion, $v=\beta/m_Q$, with $\beta$ the Gaussian
width and $m_Q$ the constituent mass.
This parameter is distinguished from the state-dependent
root-mean-square constituent velocity.
Current-component independence is preserved through
$\mathcal O(v^2)$.
We also evaluate the $\mathcal O(v^4)$ corrections numerically
to assess the size of higher-order effects.
The analytic results reveal how the radial overlap controls the
leading $D$-term.
The pure Gaussian $1S$ contribution begins at $\mathcal O(v^2)$,
whereas the $2S$ basis state retains an $\mathcal O(v^0)$ contribution.
The interference first contributes at $\mathcal O(v^2)$.
Consequently, a nonzero $2S$ admixture at fixed mixing angle leaves
an $\mathcal O(v^0)$ $D$-term in the physical ground state.
The suppression of the pure $1S$ contribution is specific to the
Gaussian overlap in the present constituent-current prescription,
rather than a general property of nodeless ground-state wave functions.

The paper is organized as follows.
Section~\ref{Sec:II} specifies the BT-based LFQM, and
Sec.~\ref{Sec:III} presents the GFF extraction and tests
current-component independence.
Section~\ref{Sec:NRQCD-summary} gives the analytic NR results through
$\mathcal O(v^2)$ for the two HO basis states and their interference.
Section~\ref{Sec:V} presents the numerical results and their comparison
with the NR expansion, and Sec.~\ref{Sec:VI} summarizes the findings.
{Appendices~\ref{Sec:cbar-verification} and~\ref{Sec:NRQCD} present the extraction of $\bar c$
and the detailed analytic NR derivations, respectively.}

\section{Light-front quark model in the Bakamjian--Thomas construction}
\label{Sec:II}

Our fBT-LFQM describes a $q\bar q$ meson of total momentum $P$
within a fixed two-particle Fock sector, expanded in a basis of
noninteracting, on-shell constituent quark--antiquark states.
The bound-state interaction enters through the mass operator
$\hat M=M_0+V_{q\bar q}$, where $M_0$ is the free invariant-mass
operator and $V_{q\bar q}$ describes the interaction between the
constituents~\cite{CJ97,CJ99a,CJLR15,ACJO22}.
We first specify the LF kinematics and spin coupling, then describe
the mass eigenvalue problem and the current-extraction prescription.

On the LF hypersurface $x^+=x^0+x^3=0$, seven Poincar\'e
generators are kinematic and interaction independent.
The remaining three, the LF Hamiltonian and the two transverse
rotation generators, are dynamical.~\cite{Dirac49}.

We use LF coordinates $a^\pm=a^0\pm a^3$ and
$\bm a_\perp=(a^1,a^2)$, with
$a\cdot b=\tfrac12(a^+b^-+a^-b^+)-\bm a_\perp\cdot\bm b_\perp$.
For a meson with kinematic momentum components
$(P^+,\bm P_\perp)$, the constituent momenta satisfy
$p_1^++p_2^+=P^+$ and
$\bm p_{1\perp}+\bm p_{2\perp}=\bm P_\perp$.
The internal variables
$x=p_1^+/P^+$ and
$\bm k_\perp=\bm p_{1\perp}-x\bm P_\perp$
are invariant under kinematic LF boosts. For constituent masses
$m_1$ and $m_2$, the free invariant mass is
\begin{equation}
  M_{0}^{2}=\frac{m_{1}^{2}+\bm{k}_{\perp}^{2}}{x}
    +\frac{m_{2}^{2}+\bm{k}_{\perp}^{2}}{1-x},
  \label{eq:M0}
\end{equation}
and the longitudinal motion is expressed through the relative momentum $k_z$,
\begin{equation}
  x=\frac{E_{1}+k_{z}}{E_{1}+E_{2}},\qquad
  E_{i}=\sqrt{m_{i}^{2}+\bm{k}_{\perp}^{2}+k_{z}^{2}},
  \label{eq:kz}
\end{equation}
so that $M_0=E_1+E_2$ and
$\vec k=(\bm k_\perp,k_z)$ is the relative three-momentum in the
rest frame of the free pair. The constituent on-shell conditions
$p_i^-=(m_i^2+\bm p_{i\perp}^2)/p_i^+$ give
\begin{equation}
  P^{-}_{\rm free}=p_{1}^{-}+p_{2}^{-}
   =\frac{M_{0}^{2}+\bm{P}_{\perp}^{2}}{P^{+}},
  \label{eq:Pminus-free}
\end{equation}
which is the energy of the free basis state rather than that
of the interacting meson.

The spin coupling is constructed by relating the canonical
spins in the pair rest frame to LF helicities through the
momentum-dependent Melosh rotation~\cite{Melosh}. For a constituent
of mass $m$ and internal variables $(x,\bm k_\perp)$, it is
\begin{equation}
  R_{M}(x,\bm{k}_{\perp},m)=
  \frac{m+x M_{0}-i\,\bm{\sigma}\!\cdot\!
        (\hat{\mathbf{z}}\times\bm{k}_{\perp})}
       {\sqrt{(m+x M_{0})^{2}+\bm{k}_{\perp}^{2}}}\,,
  \label{eq:melosh}
\end{equation}
where $\bm\sigma$ are the Pauli matrices and
$\hat{\mathbf z}$ is the longitudinal unit vector.
The rotation satisfies $R_M^\dagger R_M=1$ and tends to the
identity for $\bm k_\perp\to0$.
Coupling the rotated spins with Clebsch--Gordan coefficients gives
the spin--orbit wave function
$\mathcal R^{JJ_z}_{\lambda_1\lambda_2}(x,\bm k_\perp)$,
where $\lambda_1$ and $\lambda_2$ are the quark and antiquark
LF helicities.

For the pseudoscalar quarkonia considered here, $J=0$ and
$m_1=m_2=m\equiv m_Q$, where $Q=c,b$. The equal-mass relations are
$M_0^2=(m^2+\bm k_\perp^2)/[x(1-x)]$,
$k_z=(x-\tfrac12)M_0$,
$\partial k_z/\partial x=M_0/[4x(1-x)]$, and
$\vec k^{\,2}=M_0^2/4-m^2$.
The pseudoscalar spin--orbit function takes the covariant
form~\cite{SLF2,SLF3}
\begin{equation}
  \mathcal{R}_{\lambda_{1}\lambda_{2}}^{00}
  =\frac{\bar{u}_{\lambda_{1}}(p_{1})\,\gamma_{5}\,
         v_{\lambda_{2}}(p_{2})}{\sqrt{2}\,M_{0}}\,,
  \label{eq:spinorbit}
\end{equation}
or, equivalently, the matrix form
\begin{align}\label{eq:SOmatrix}
  {\cal R}^{00}_{\lambda_1\lambda_2}
  =\frac{1}{\sqrt{2x(1-x)}\,M_0}
  \left(
  \begin{array}{cc}
    -k^{(-)} & m \\
    -m & -k^{(+)}
  \end{array}
  \right),
\end{align}
where $k^{(\pm)}=k_x\pm ik_y$, and the rows and columns are
ordered by $\lambda_1=\uparrow,\downarrow$ and
$\lambda_2=\uparrow,\downarrow$, respectively.
The spin--orbit normalization is
$\sum_{\lambda_1\lambda_2}
|\mathcal R^{00}_{\lambda_1\lambda_2}|^2=1$.

The internal LFWF factorizes into spin--orbit and radial parts,
\begin{equation}
  \Psi_{\lambda_1\lambda_2,nS}^{00}(x,\bm{k}_\perp)
  =\Phi_{nS}(x,\bm{k}_\perp)\,
   \mathcal{R}_{\lambda_1\lambda_2}^{00}(x,\bm{k}_\perp).
  \label{eq:LFWF}
\end{equation}
The radial ansatz of Ref.~\cite{ACJO22} is
\begin{eqnarray}\label{eq:HOWF1}
  \begin{pmatrix}
    \Phi_{1S} \\
    \Phi_{2S}
  \end{pmatrix}
  =
  \begin{pmatrix}
    \cos\theta & \sin\theta \\
    -\sin\theta & \cos\theta
  \end{pmatrix}
  \begin{pmatrix}
    \phi_{1S} \\
    \phi_{2S}
  \end{pmatrix},
\end{eqnarray}
with the two HO basis functions
\begin{eqnarray}\label{eq:HOWF2}
  \phi_{1S}(x,\bm{k}_\perp)
  &=&
  \frac{4\pi^{3/4}}{\beta^{3/2}}
  \sqrt{\frac{\partial k_z}{\partial x}}\,
  e^{-\vec{k}^{\,2}/(2\beta^2)},
  \nonumber\\
  \phi_{2S}(x,\bm{k}_\perp)
  &=&
  \frac{4\pi^{3/4}}{\sqrt{6}\,\beta^{7/2}}
  \sqrt{\frac{\partial k_z}{\partial x}}\,
  \left(2\vec{k}^{\,2}-3\beta^2\right)
  e^{-\vec{k}^{\,2}/(2\beta^2)}.
\end{eqnarray}
For a given quark flavor, the two HO basis states share a common
Gaussian width $\beta$.
The square-root Jacobian ensures their orthonormality under the
LF integration measure,
\begin{equation}
\int_0^1 dx \int \frac{d^2\bm{k}_\perp}{2(2\pi)^3}\,
\phi_{nS}^{*}(x,\bm{k}_\perp)
\phi_{n'S}(x,\bm{k}_\perp)
=
\delta_{nn'},
\qquad n,n'=1,2.
\end{equation}
The orthogonal mixing matrix preserves this orthonormality.
Together with the spin--orbit normalization, it also ensures the
normalization and orthogonality of the full LFWFs after summing
over the constituent helicities.
The spectroscopy analysis of Ref.~\cite{ACJO22} found this mixing
important for the simultaneous description of the masses and decay
constants of the $1S$ and $2S$ heavy mesons.

The interaction $V_{q\bar q}$ is independent of the total momentum
and invariant under the kinematic subgroup.
For an interaction satisfying the BT constraints, the kinematic
generators retain their free form, while the dynamical generators
are constructed consistently from $\hat M$ so that the Poincar\'e
algebra is preserved on the two-particle
space~\cite{BT53,Poly10,CCKP}.
In this interacting representation, $\hat M$ is the mass Casimir
and commutes with all Poincar\'e generators.
In particular, the LF Hamiltonian is
\begin{equation}
  P^-_{\rm int}
  =\frac{\hat M^2+\bm P_\perp^2}{P^+},
  \label{eq:Pminus-int}
\end{equation}
and the bound states satisfy
\begin{equation}
  \hat M|\Psi\rangle=M_{q\bar q}\,|\Psi\rangle,
  \label{eq:eigen}
\end{equation}
where $M_{q\bar q}$ is the physical meson mass and
the internal LFWF is the corresponding eigenfunction.%
\footnote{In Refs.~\cite{CJ97,CJ99a,CJLR15,ACJO22}, the mass
operator was denoted $H_{q\bar q}=M_0+V_{q\bar q}$.
Here $\hat M$ distinguishes the operator from its eigenvalue
$M_{q\bar q}$ and the free invariant mass
$M_0(x,\bm k_\perp)$.}
In practice, this eigenfunction is approximated by the HO ansatz
of Eqs.~\eqref{eq:HOWF1} and~\eqref{eq:HOWF2}.
At fixed $P^+$ and $\bm P_\perp$, Eq.~\eqref{eq:Pminus-int}
relates this internal eigenvalue problem to that of the
LF Hamiltonian.

We adopt the confining-plus-hyperfine
potential~\cite{ACJO22}
\begin{equation}
  V_{q\bar q}
  =a+br-\frac{4\alpha_s}{3r}+V_{\rm hyp},
  \label{eq:potential}
\end{equation}
where $r$ is the interquark separation and $V_{\rm hyp}$ is the
spin--spin hyperfine interaction.
The potential combines linear confinement with a Coulomb-like
one-gluon-exchange term.
In the spectroscopic analysis, the Gaussian width $\beta$ is determined
variationally from the kinetic and central-potential terms, { typically} with the
hyperfine interaction treated perturbatively.
The mixing angle $\theta$ is constrained by the mass splittings and
decay-constant hierarchy of the $1S$ and $2S$ heavy mesons.
We adopt the resulting parameter set with $\theta=12^\circ$ and use
the radial wave functions in the GFF calculation without further tuning.

The mass operator determines the bound-state dynamics but
does not by itself specify a unique interacting current operator.
For current matrix elements, we use the integrand-level fBT
prescription introduced in Sec.~\ref{Sec:I}.
For a schematic single-form-factor decomposition
$\langle P'|\mathcal O^\mu|P\rangle
=\mathcal P^\mu\mathcal F$, where $\mathcal P^\mu$ is the
kinematic Lorentz structure, we write
\begin{equation}\label{eq:self-consistent}
  \mathcal F^{(\mu)}
  =\left\langle P'\left|
  \frac{\mathcal O^\mu}{\mathcal P^\mu}
  \right|P\right\rangle_{\rm BT}.
\end{equation}
Here $\mu$ labels the current component used for the extraction
and is not summed.
The constituent matrix element and Lorentz prefactor are evaluated
with $M\to M_0$ and $M'\to M'_0$ inside the overlap integral.
The masses $M$ and $M'$ refer to the initial and final physical
mesons, whereas $M_0$ and $M'_0$ are the corresponding
free-constituent invariant masses.
This notation denotes an integrand-level ratio rather than the
division of an integrated matrix element by an external prefactor.
The replacements apply to the current-extraction integrand and
do not alter the bound-state mass eigenvalue problem.
Section~\ref{Sec:III} specifies the prescription for the EMT
and its two GFFs.

\section{Gravitational form factors}\label{Sec:III}

We consider elastic matrix elements of the constituent-quark EMT
between pseudoscalar quarkonium states with momenta $P$ and $P'$
and equal physical masses $M=M'$.
These physical masses are distinguished from the free-constituent
invariant masses $M_0$ and $M'_0$ entering the fBT prescription.
The symmetric constituent-quark EMT operator is
\begin{align}\label{eq:EMT}
  \hat{T}_q^{\mu\nu}(x)
  =\frac{i}{4}\,\bar\psi_q(x)\!\left(
    \gamma^\mu\overleftrightarrow{\partial}{}^\nu
    +\gamma^\nu\overleftrightarrow{\partial}{}^\mu
  \right)\!\psi_q(x),
\end{align}
where
$\overleftrightarrow{\partial}{}^\mu
=\overrightarrow{\partial}{}^\mu
-\overleftarrow{\partial}{}^\mu$.
Within the valence $Q\bar Q$ truncation adopted here, this operator
is constructed with the ordinary derivative rather than the
covariant one.
The binding dynamics enters through the mass operator $\hat M$,
rather than through an explicit gluon EMT.

We first evaluate the contribution in which the EMT couples to
the quark while the antiquark remains a spectator.
Denoting this contribution by the subscript ``quark'' on the
matrix element, we write
\begin{align}\label{eq:GFFdecomp}
  \langle T^{\mu\nu}_q\rangle
  &\equiv
  \langle P'|\hat{T}^{\mu\nu}_q(0)|P\rangle_{\mathrm{quark}}
  \nonumber\\
  &=
  \Gamma_A^{\mu\nu}\,A_q(t)
  +\Gamma_D^{\mu\nu}\,D_q(t)
  +2M^2\,\bar c_q(t)\,g^{\mu\nu},
\end{align}
where $\Delta=P'-P$ is the momentum transfer,
$t=\Delta^2<0$, and $\bar P=\tfrac12(P+P')$ is the average
meson momentum.
The associated Lorentz structures are
\begin{align}\label{eq:LorentzStruct}
  \Gamma_A^{\mu\nu}
  &=
  2\left[
    \bar P^\mu\bar P^\nu
    -\frac{\bar P\cdot\Delta}{\Delta^2}
    \left(
      \Delta^\mu\bar P^\nu
      +\Delta^\nu\bar P^\mu
      -\bar P\cdot\Delta\,g^{\mu\nu}
    \right)
  \right],
  \nonumber\\
  \Gamma_D^{\mu\nu}
  &=
  \frac12
  \left(
    \Delta^\mu\Delta^\nu-g^{\mu\nu}\Delta^2
  \right),
\end{align}
with $\bar P\cdot\Delta=(M'^2-M^2)/2$.
For the physical elastic states with $M=M'$, one has
$\bar P\cdot\Delta=0$ and
$\Gamma_A^{\mu\nu}=2\bar P^\mu\bar P^\nu$.
In the fBT prescription, however, the Lorentz structures in
Eq.~\eqref{eq:LorentzStruct} are evaluated in their unreduced form
inside the overlap integral, with $M\to M_0$ and $M'\to M'_0$.
This gives
\begin{equation}\label{eq:PbarDeltaBT}
  \left[\bar P\cdot\Delta\right]_{\mathrm{BT}}
  =\frac{M_0'^2-M_0^2}{2}.
\end{equation}
Since $M_0$ and $M'_0$ depend on the initial and recoil-shifted
internal momenta, respectively, this quantity is generally nonzero
at finite momentum transfer.
The terms involving $\bar P\cdot\Delta$ in
Eq.~\eqref{eq:LorentzStruct} must therefore be retained in the
fBT prefactors even for elastic scattering.
The subscript $\mathrm{BT}$ denotes this integrand-level evaluation
and does not imply a change in the physical elastic kinematics.

{The form factor $\bar c_q(t)$ allows for possible
nonconservation of the active-quark contribution.
Applying the fBT prescription to its prefactor gives
$\Gamma_{\bar c}^{\mu\nu}=2M_0^2g^{\mu\nu}$.
The independent extraction in Appendix~A yields
$\bar c_q(t)=0$, so the following projections determine
$A_q(t)$ and $D_q(t)$ without an additional metric term.}

Since the heavy quarkonia considered here are equal-mass $Q\bar Q$
states with identical quark and antiquark contributions, the total
valence GFFs are
\begin{align}\label{eq:GFFtotal}
  A_{Q\bar Q}(t)=2A_q(t),\qquad
  D_{Q\bar Q}(t)=2D_q(t).
\end{align}
{Combining this symmetry with the vanishing $\bar c_q(t)$
established in Appendix~\ref{Sec:cbar-verification} gives
$\bar c_{Q\bar Q}(t)=2\bar c_q(t)=0$.}

We work in the $\Delta^+=0$ frame with
$\bm P_\perp=\bm0_\perp$, where the initial- and final-state
four-momenta are
\begin{align}\label{eq:momenta}
  P
  &=
  \left(
    P^+,\,\frac{M^2}{P^+},\,\bm0_\perp
  \right),
  \nonumber\\
  P'
  &=
  \left(
    P^+,\,\frac{M'^2+\bm\Delta_\perp^2}{P^+},
    \,\bm\Delta_\perp
  \right).
\end{align}
{
For the computation in the spacelike region, $t = \Delta^2 < 0$, this frame is typically useful to find the physical observable staying away from the components prone to the LF zero-mode issue.
For a schematic single-form-factor decomposition
$\langle P'|\mathcal O^\mu|P\rangle
=\mathcal P^\mu\mathcal F$ associated with Eq.(\ref{eq:self-consistent}),
{the physical observable $\mathcal F$ can be extracted
from the plus component when $\mathcal P^+\neq0$,}
without involving $\langle P'|\mathcal O^-|P\rangle$, notorious for its involvement of the LF zero-modes. By the same token, for this work, it is sufficient to compute $\langle T^{++}_q\rangle$,
$\langle T^{+-}_q\rangle$, $\langle T^{+i}_q\rangle$, $\langle T^{-i}_q\rangle$ and $\langle T^{ij}_q\rangle$
($i,j$ represent the transverse components) without involving
$\langle T^{--}_q\rangle$ to find $A_q(t)$ and $D_q(t)${.
The Lorentz structures $\Gamma_A^{\mu\nu}$ and
$\Gamma_D^{\mu\nu}$ in Eq.~\eqref{eq:LorentzStruct}
satisfy $\Delta_\mu\Gamma_A^{\mu\nu}
=\Delta_\mu\Gamma_D^{\mu\nu}=0$ by construction.}
We further employ the fBT LFQM to examine whether both $A_q(t)$ and $D_q(t)$ extracted from their available  respective components are indeed independent of the current components used for their extraction.
}

In the conventional LFQM treatment~\cite{SLF2,SLF3,CCC97},
the BT replacement is applied only to the constituent matrix
element $\langle T_q^{\mu\nu}\rangle$, while the Lorentz
prefactors retain the physical meson masses.
This partially BT-based (pBT) scheme, discussed in
Ref.~\cite{Choi:2025kaon}, differs from the uniform fBT
prescription of Eq.~\eqref{eq:self-consistent}.
In the present calculation, the forward values are obtained
from the $t\to0$ limit of the complete extraction expressions,
with all cancellations among their kinematic factors retained.

The incoming and outgoing on-shell constituent momenta are denoted
by $p_i$ and $p'_i$, with $i=1$ for the active quark and $i=2$
for the spectator antiquark.
Their plus and transverse components are parametrized as
\begin{align}\label{eq:LFmomenta}
  & p_i^+=p_i^{\prime +}=x_iP^+,
  \nonumber\\
  & \bm p_{i\perp}=x_i\bm P_\perp+\bm k_{i\perp},\qquad
    \bm p'_{i\perp}=x_i\bm P'_\perp+\bm k'_{i\perp},
\end{align}
where $x_1=x$, $x_2=1-x$,
$\bm k_{1\perp}=-\bm k_{2\perp}=\bm k_\perp$, and
$\bm k'_{1\perp}=-\bm k'_{2\perp}=\bm k'_\perp$.
The spectator condition
$\bm p'_{2\perp}=\bm p_{2\perp}$ then gives
$\bm k'_\perp =\bm k_\perp+(1-x)\bm\Delta_\perp$.
In the frame of Eq.~\eqref{eq:momenta}, the active-quark
transverse momenta are instead
$\bm p_{1\perp}=\bm k_\perp$ and $\bm p'_{1\perp}=\bm k_\perp+\bm\Delta_\perp$.
The outgoing spinor is therefore evaluated at $p'_1$, with
$p_1^{\prime +}=xP^+$ and
$p_1^{\prime -}=\frac{m^2+(\bm k_\perp+\bm\Delta_\perp)^2}{xP^+}$,
whereas the final-state LFWF is evaluated at the relative
momentum $\bm k'_\perp$.

Using the on-shell constituent basis of the BT construction,
we evaluate the matrix element as an overlap convolution of
the initial- and final-state LFWFs~\cite{Choi:2025},
\begin{align}\label{eq:convolution}
  \langle T_q^{\mu\nu}\rangle_{\rm BT}
  =
  \int [d^3\bm p_1]\,
  \Phi'^{*}(x,\bm k'_\perp)\,
  \Phi(x,\bm k_\perp)\,
  S_q^{\mu\nu}.
\end{align}
The integration measure can be written as
$[d^3\bm p_1]=P^+\frac{{\rm d}x\,{\rm d}^2\bm k_\perp}{2(2\pi)^3}$.
The spin trace for the tensor current is
\begin{align}\label{eq:spintrace}
  S_q^{\mu\nu}
  &=
  \sum_{\lambda_1,\lambda'_1,\lambda_2}
  \mathcal R'^{*}_{\lambda'_1\lambda_2}\,
  \frac{\bar u_{\lambda'_1}(p'_1)}
       {\sqrt{p_1^{\prime +}}}
  \left[
    \frac14
    \left(
      p_s^\mu\gamma^\nu+p_s^\nu\gamma^\mu
    \right)
  \right]
  \frac{u_{\lambda_1}(p_1)}{\sqrt{p_1^+}}\,
  \mathcal R_{\lambda_1\lambda_2}
  \nonumber\\
  &\equiv
  \sum_{\lambda_1,\lambda'_1,\lambda_2}
  \mathcal R'^{*}_{\lambda'_1\lambda_2}\,
  U^{\mu\nu}_{\lambda_1\to\lambda'_1}(p_1\to p'_1)\,
  \mathcal R_{\lambda_1\lambda_2},
\end{align}
where $p_s=p_1+p'_1$,
$\mathcal R_{\lambda_1\lambda_2}
=\mathcal R^{00}_{\lambda_1\lambda_2}(x,\bm k_\perp)$, and
$\mathcal R'_{\lambda'_1\lambda_2}
=\mathcal R^{00}_{\lambda'_1\lambda_2}(x,\bm k'_\perp)$
are the spin--orbit wave functions of
Eq.~\eqref{eq:SOmatrix}.
The spectator antiquark carries the same helicity $\lambda_2$
in the initial and final states.

Evaluating the Dirac bilinears in the helicity basis~\cite{BD80}
gives the matrix elements
$U^{\mu\nu}_{\lambda_1\to\lambda'_1}$
listed in Table~\ref{tab:ME}.
The helicity sums in Eq.~\eqref{eq:spintrace} then yield
the spin traces $S_q^{\mu\nu}$, which are collected in
Table~\ref{tab:trace} together with the Lorentz structures
$\Gamma_A^{\mu\nu}$ and $\Gamma_D^{\mu\nu}$.
{At fixed internal variables, the combinations of spin traces
and Lorentz prefactors used for GFF extraction, including
the $A_q$ subtraction where required, carry an overall factor
$1/P^+$ before the measure is included.}
This cancels the factor $P^+$ in the integration measure,
leaving $A_q(t)$ and $D_q(t)$ independent of $P^+$,
as required by kinematic longitudinal-boost invariance.

{Throughout, repeated transverse indices $i,j=x,y$
denote Euclidean sums.}
Contracting the transverse Lorentz structures with
$\Delta_i\Delta_j$ gives
\begin{align}\label{eq:transverse}
  \Gamma_D^{ij}\Delta_i\Delta_j
  &=0,
  \qquad
  \Gamma_A^{ij}\Delta_i\Delta_j
  =\frac12
  \bigl(M_-^2+\bm{\Delta}_\perp^2\bigr)^2,
\end{align}
where $M_-^2=M_0'^2-M_0^2$ denotes the difference between
the squared final- and initial-state free-constituent invariant masses.
{With $\bar c_q(t)=0$, the first identity and
$\Gamma_D^{++}=\Gamma_D^{+i}=0$ in Table~\ref{tab:trace} allow $A_q(t)$ to be extracted
from the $(++)$, $(+i)$, and $(ij)$ channels.}
The $(+-)$, $(-i)$, and $(ii)$ channels provide three
extractions of $D_q(t)$ after subtracting the $A_q$ contribution
at the integrand level.
Here $(+i)$ and $(-i)$ denote contractions with $\Delta_i$,
$(ij)$ denotes the double contraction with $\Delta_i\Delta_j$,
and $(ii)$ denotes the transverse trace.

The terms proportional to $e^i$ in Table~\ref{tab:trace}
arise from products of the imaginary parts of the helicity
bilinears and spin--orbit overlaps, yielding real contributions
to the spin traces.
Since $\Delta_i e^i=0$, these terms vanish in the contracted
$(+i)$, $(-i)$, and $(ij)$ channels.
They nevertheless contribute to the transverse trace,
\be
  \left.S_q^{ii}\right|_e
  =
  \frac{(1-x)\chi^2}
     {p_1^+\sqrt{m^2+\bm k_\perp^2}
    \sqrt{m^2+\bm k_\perp^{\prime 2}}},
\ee
where $\left.S_q^{ii}\right|_e$ denotes the part of the trace
arising from the $e^i$ terms.
This contribution must therefore be retained in the $(ii)$
extraction of $D_q(t)$.

%%%%%%%%%%%%% Table I
\begin{table*}[t]
    \caption{\label{tab:ME}
    Matrix elements of $U^{\mu\nu}_{\lambda_1\to \lambda'_1}(p_1\to p'_1)$, as defined in Eq.~\eqref{eq:spintrace}. Here,
     $p_{(\pm)}=p^1\pm i p^2$ and $\delta_{(\pm)}^i=\delta^{i1}\pm i\delta^{i2}$.
     Since \( p^+_1 = p^{\prime +}_1 \) and $\delta p_{(\pm)}=(p_{1(\pm)} - p'_{1(\pm)})$ hold in this framework, the results are expressed accordingly.}
    \centering
    \setlength{\tabcolsep}{6pt}%           
    %\small%                                 
    \begin{tabular}{@{}ccc@{}}%             
    \hline \hline \\[-1ex]
    Matrix & \multicolumn{2}{c}{Helicity $\lambda_1 \to \lambda'_1$} \\
    elements  &  &  \\
    $U^{\mu\nu}$ &
     $ \left\{ \begin{array}{c}
    \uparrow \;\to\; \uparrow \\
    \downarrow \;\to\;  \downarrow
   \end{array}  \right\} $
    & $ \left\{ \begin{array}{c}
    \uparrow \;\to\; \downarrow \\
    \downarrow \;\to\;  \uparrow
   \end{array}  \right\} $  \\[3.5ex]
    \hline \\
    $U^{++} $ & $2p_1^+$ & 0 \\[3.5ex]
    $U^{+-}$ & $\frac{1}{2}
    \left(p_1 + p'_1\right)^- + \frac{1}{p_1^+}\left(m^2+ p_{1(\pm)} p'_{1(\mp)} \right)$ &
    $ \pm \frac{m}{p_1^+ }\delta p_{(\pm)} $\\[3.5ex]
    $U^{+i}$ &
      $\frac{1}{2}\left( p_{1}+ p'_{1}\right)^{i} + \frac{1}{2}
      \left( p_{1(\pm)} \delta^i_{(\mp)} + p'_{1(\mp)} \delta^i_{(\pm)} \right)$ & 0
      \\[3.5ex]
    $U^{-i}$ &
    $\frac{( p_{1}+ p'_{1})^i}{2(p_1^+)^{2}} \left(  m^2+ p_{1(\pm)} p'_{1(\mp)} \right)
     +
    \frac{(p_{1} +p'_{1})^-}{4p_1^+} \left(  p_{1(\pm)} \delta^i_{(\mp)} + p'_{1(\mp)} \delta^i_{(\pm)} \right)
    $ &
    $\pm\frac{ m( p_{1}+ p'_{1})^i}{2(p_1^+)^{2}} \delta p_{(\pm)}  $
    \\[3.5ex]
    $U^{ij}$ &
    $\frac{( p_{1}+ p'_{1})^i}{4 p_1^+}
    \left( p_{1(\pm)} \delta^j_{(\mp)} + p'_{1(\mp)} \delta^j_{(\pm)}  \right)
    +\frac{( p_{1}+ p'_{1})^j}{4 p_1^+}
    \left( p_{1(\pm)} \delta^i_{(\mp)} + p'_{1(\mp)} \delta^i_{(\pm)}  \right) $  & 0
    \\[3.5ex]
      \hline\hline
    \end{tabular}
\end{table*}
%%%%%%%%%%%%%%%%%%%%%%%%%%

%%%%%%%%%%%%% Table II    
\begin{table*}
    \caption{\label{tab:trace}%
Spin traces $S^{\mu\nu}_q$ and Lorentz structures $\Gamma_A^{\mu\nu}$,
$\Gamma_D^{\mu\nu}$ for the tensor current. Here
$S_c=(m^2+\vk\cdot\vkp)(\vk^2+\vk\cdot\bm{\Delta}_{\perp}+m^2)
+(1-x)(\vk\times\bm{\Delta}_{\perp})^2+(1-x)m^2\bm{\Delta}_{\perp}^2$,
$M^2_{\pm}=M_{0}^{\prime 2}\pm M_0^2$,
{$p_s^i=2k^i+\Delta^i$, $e^i=\epsilon^{ij}\Delta^j$,
$\chi=\vk\times\bm{\Delta}_{\perp}$, and $\epsilon^{xy}=-\epsilon^{yx}=+1$}.}
    \begin{tabular}{c@{\hskip 0.5cm}c@{\hskip 0.5cm}c@{\hskip 0.5cm}c}
      \hline\hline \\[-1.5ex]
    $(\mu\nu)$ & $S^{\mu\nu}_q$ & $\Gamma_A^{\mu\nu}$ & $\Gamma_D^{\mu\nu}$ \\[1.5ex]
    \hline \\[-1.5ex]
    $(++)$ & $\frac{2p_1^+(m^2+\vk\cdot\vkp)}{\sqrt{m^2+\vk^2}\sqrt{m^2+\vk'^2}}$
    & $2(\bar{P}^+)^2$ & 0 \\[3.5ex]
    $(+-)$ & $\frac{p^+_1 p_s^-(m^2+\vk\cdot\vk')+2S_c}{2p^+_1\sqrt{m^2+\vk^2}\sqrt{m^2+\vk'^2}}$
    & $M^2_+ + M^2_- + \bm{\Delta}_{\perp}^2$ & $\bm{\Delta}_{\perp}^2$ \\[3.5ex]
    $(+i)$ & $\frac{p_s^i(m^2+\vk\cdot\vkp){+\tfrac{1}{2}(1-x)\chi e^i}}{\sqrt{m^2+\vk^2}\sqrt{m^2+\vk'^2}}$
    & $\bar{P}^+\Delta^i\left(1+\frac{M^2_-}{\bm{\Delta}_{\perp}^2}\right)$ & 0 \\[3.5ex]
    $(-i)$ & $\frac{p_s^i\left[\frac{1}{2}p^+_1 p_s^-(m^2+\vk\cdot\vkp)+S_c\right]
    {+\tfrac{1}{2}(1-x)p^+_1 p_s^-\chi e^i}}{2(p^+_1)^{2}\sqrt{m^2+\vk^2}\sqrt{m^2+\vk'^2}}$
    & $\frac{\Delta^i(\bm{\Delta}_{\perp}^2+M^2_-)(\bm{\Delta}_{\perp}^2+M^2_++M^2_-)}{2\bm{\Delta}_{\perp}^2\bar{P}^+}$
    & $\frac{\Delta^i(\bm{\Delta}_{\perp}^2+M^2_-)}{2\bar{P}^+}$ \\[3.5ex]
    $(ij)$ & $\frac{p_s^ip_s^j(m^2+\vk\cdot\vkp)
    {+\tfrac{1}{2}(1-x)\chi(p_s^i e^j + p_s^j e^i)}}
    {2p^+_1\sqrt{m^2+\vk^2}\sqrt{m^2+\vk'^2}}$
    & $\frac{1}{2}\left[\Delta^i\Delta^j\left(1+\frac{2M^2_-}{\bm{\Delta}_{\perp}^2}\right)-g^{ij}\frac{M^4_-}{\bm{\Delta}_{\perp}^2}\right]$
    & $\frac{1}{2}(\Delta^i\Delta^j+g^{ij}\bm{\Delta}_{\perp}^{2})$ \\[3.5ex]
    \hline\hline
    \end{tabular}
\end{table*}

Following the fBT prescription, we express the three extractions
of $A_q(t)$ as convolutions of the radial wave functions with
channel-dependent kernels,
\begin{equation}\label{eq:Aextract}
  A_q^{(\alpha)}(t)
  =
  \int_0^1 dx
  \int \frac{d^2\bm k_\perp}{2(2\pi)^3}\,
  \Phi'^{*}(x,\bm k'_\perp)\,
  \Phi(x,\bm k_\perp)\,
  \mathcal{K}_A^{(\alpha)}(x,\bm k_\perp,\bm\Delta_\perp),
  \qquad
  \alpha=++,+i,ij.
\end{equation}
The dimensionless extraction kernels are
\begin{align}\label{eq:Akernels}
  \mathcal{K}_A^{(++)}
  &=
  P^+\frac{S_q^{++}}{\Gamma_A^{++}},
  \nonumber\\
  \mathcal{K}_A^{(+i)}
  &=
  P^+\frac{S_q^{+i}\Delta_i}
      {\Gamma_A^{+i}\Delta_i},
  \nonumber\\
  \mathcal{K}_A^{(ij)}
  &=
  P^+\frac{S_q^{ij}\Delta_i\Delta_j}
    {\Gamma_A^{ij}\Delta_i\Delta_j}.
\end{align}
Here the factor $P^+$ from the original momentum measure
$[d^3\bm p_1]$ has been absorbed into the kernels.
The spin traces and Lorentz prefactors in each ratio are
evaluated with the fBT prescription inside the overlap integral.

For equal constituent masses,
$M_0^2=(m_Q^2+\bm k_\perp^2)/[x(1-x)]$
and the spectator shift give
\be
  M_-^2+\bm\Delta_\perp^2
  =
  \frac{2\bm k_\perp\cdot\bm\Delta_\perp
     +\bm\Delta_\perp^2}{x}.
\ee
Using this relation, the common kinematic factors cancel
between the numerators and denominators of the $(+i)$ and
$(ij)$ kernels.
The three reduced kernels therefore coincide pointwise,
\begin{equation}\label{eq:A-universal}
  \mathcal{K}_A^{(++)}
  =
  \mathcal{K}_A^{(+i)}
  =
  \mathcal{K}_A^{(ij)}
  \equiv \mathcal{K}_A
  =
  x\,
  \frac{m_Q^2+\bm k_\perp\cdot\bm k'_\perp}
       {\sqrt{m_Q^2+\bm k_\perp^2}
        \sqrt{m_Q^2+\bm k_\perp^{\prime 2}}}.
\end{equation}
%{At zeros of the common kinematic factors, the reduced
%expression defines the continuous extension of the contracted
%ratios in Eq.~\eqref{eq:Akernels}.}
The kernel identity is independent of the radial wave functions.
Convolution with the same radial overlap therefore gives
\be
  A_q(t)
  =
  A_q^{(++)}(t)
  =
  A_q^{(+i)}(t)
  =
  A_q^{(ij)}(t).
\ee
Keeping $\Gamma_A^{\mu\nu}$ at the physical elastic masses
removes the invariant-mass-difference terms from the contracted
prefactors and spoils this kernel identity.

For the $(+-)$, $(-i)$, and $(ii)$ channels, we subtract the
$A_q$ contribution at the integrand level using the common
kernel \revCbar{$\mathcal K_A$} of Eq.~\eqref{eq:A-universal}.
The three extractions of $D_q(t)$ are written as
\begin{equation}\label{eq:Dextract}
  D_q^{(\alpha)}(t)
  =
  \int_0^1 dx
  \int \frac{d^2\bm k_\perp}{2(2\pi)^3}\,
  \Phi'^{*}(x,\bm k'_\perp)\,
  \Phi(x,\bm k_\perp)\,
  \mathcal{K}_D^{(\alpha)}(x,\bm k_\perp,\bm\Delta_\perp),
  \qquad
  \alpha=+-,-i,ii.
\end{equation}
The corresponding dimensionless kernels are
\begin{align}\label{eq:Dkernels}
  \mathcal{K}_D^{(+-)}
  &=
  \frac{P^+S_q^{+-}-\Gamma_A^{+-}\mathcal{K}_A}
       {\Gamma_D^{+-}},
  \nonumber\\
  \mathcal{K}_D^{(-i)}
  &=
  \frac{P^+S_q^{-i}\Delta_i
      -\bigl(\Gamma_A^{-i}\Delta_i\bigr)\mathcal{K}_A}
     {\Gamma_D^{-i}\Delta_i},
  \nonumber\\
  \mathcal{K}_D^{(ii)}
  &=
  \frac{P^+S_q^{ii}-\Gamma_A^{ii}\mathcal{K}_A}
       {\Gamma_D^{ii}}.
\end{align}
All spin traces and Lorentz prefactors are evaluated with
the fBT prescription inside the overlap integral.

{The reduced $(+-)$ and momentum-transfer-contracted $(-i)$
kernels coincide pointwise, as shown in Appendix~\ref{Sec:cbar-verification}.
The transverse-trace kernel differs pointwise, but its convolution
with the adopted radial $S$-wave overlaps agrees with the other
two extractions within numerical accuracy. Thus,}
\be\label{eq:D-component-independence}
  D_q(t)  =  D_q^{(+-)}(t)  =  D_q^{(-i)}(t)  =  D_q^{(ii)}(t).
\ee
%The subtraction is unchanged if { $\mathcal{K}_A$} is evaluated using
%any of the three expressions in Eq.~\eqref{eq:Akernels}.

For the radius definitions below, we use the total valence
form factors $A(t)\equiv A_{Q\bar Q}(t)$ and
$D(t)\equiv D_{Q\bar Q}(t)$.
For a form factor $F(t)$ with $F(0)\neq0$, we define the
conventional three-dimensional mean-square slope radius by
\begin{equation}\label{eq:radius}
  \langle r_F^2\rangle
  =
  \frac{6}{F(0)}
  \left.\frac{dF(t)}{dt}\right|_{t=0}.
\end{equation}
The choice $F=A$, with $A(0)=1$, gives the mass radius
$\sqrt{\langle r_A^2\rangle}$.
The choice $F=D$ defines the $D$-term slope radius
$\sqrt{\langle r_D^2\rangle}$.

The mechanical radius is instead defined from the
three-dimensional normal-force distribution~\cite{Polyakov:2018zvc},
\begin{equation}\label{eq:rmech}
  \langle r^2\rangle_{\mathrm{mech}}
  =
  \frac{\displaystyle
     \int d^3r\,r^2
     \left[\frac23s(r)+p(r)\right]}
       {\displaystyle
     \int d^3r\,
     \left[\frac23s(r)+p(r)\right]}
  =
  \frac{6D(0)}
  {\displaystyle\int_{-\infty}^{0}dt\,D(t)},
\end{equation}
where $p(r)$ and $s(r)$ are the three-dimensional pressure
and shear-force distributions, respectively.
Unlike the $D$-term slope radius, this definition depends on
$D(t)$ over the full spacelike region rather than its slope
at $t=0$.
Under the local stability condition $\tfrac23s(r)+p(r)>0$,
both coordinate-space integrals in Eq.~\eqref{eq:rmech}
are positive, provided they converge.
These three-dimensional radii are distinct from the transverse
density moments introduced in Sec.~\ref{Sec:densities}.

\section{Nonrelativistic expansion through next-to-leading order}
\label{Sec:NRQCD-summary}

We present the analytic expansions through $\mathcal O(v^2)$
for the diagonal contributions of the two HO basis states and
their interference, together with their implications for the
physical $1S$--$2S$ mixtures.
Detailed coefficient expansions and Gaussian integrations are
given in Appendix~\ref{Sec:NRQCD}.

%\subsection{Expansion scheme and power counting}
%\label{Sec:NR-power-counting}

We expand in $v=\beta/m_Q\ll1$ at fixed
$|\bm\Delta_\perp|/\beta$ and $\vec k/\beta$, and define
$\rho=\bm\Delta_\perp^2/\beta^2=-t/\beta^2$.
For background on the nonrelativistic description of heavy
quarkonium, see
Refs.~\cite{Quigg:1979vr,Bodwin:1994jh,Caswell:1985ui,
Brambilla:2004jw,Brambilla:2010cs}.
Throughout this section, NLO denotes the common truncation
through $\mathcal O(v^2)$.
When the $\mathcal O(v^0)$ contribution vanishes, the retained
$\mathcal O(v^2)$ term is the first nonzero result.

Changing variables from $x$ to $k_z$ exactly, we write
$F_q=A_q$ or $D_q$ as
\begin{align}
  F_q(t)
  &=
  \int\!\frac{dk_z\,d^2\bm k_\perp}{2(2\pi)^3}\,
  \mathcal W(\vec k,\bm\Delta_\perp;v)\,
  \mathcal{K}_F(\vec k,\bm\Delta_\perp;v),
  \label{eq:NR-KW-convolution}\\
  \mathcal W
  &\equiv
  \sqrt{\frac{M'_0}{M_0}}\,
  \widetilde\phi^{*}(\vec k\,')\,
  \widetilde\phi(\vec k),
  \label{eq:NR-KW-overlap}
\end{align}
where
$\phi=\sqrt{\partial k_z/\partial x}\,\widetilde\phi$.
The Jacobian ratio $\sqrt{M'_0/M_0}$ follows from the exact
change of measure in Eq.~\eqref{eq:NR-measure}.
The transverse spectator shift specified in Sec.~\ref{Sec:III}
and the longitudinal relation $k'_z=k_zM'_0/M_0$ are retained
before expansion, with $x=x(k_z,\bm k_\perp)$.
For the interference contribution, the radial product is
replaced by the sum of both ordered overlaps,
\[
  \widetilde\phi_{1S}^{*}(\vec k\,')\,
  \widetilde\phi_{2S}(\vec k)
  +
  \widetilde\phi_{2S}^{*}(\vec k\,')\,
  \widetilde\phi_{1S}(\vec k).
\]

The dimensionless kernels $\mathcal{K}_F^{(\alpha)}$ defined in
Eqs.~\eqref{eq:Akernels} and~\eqref{eq:Dkernels}
are evaluated with $x=x(k_z,\bm k_\perp)$ in
Eq.~\eqref{eq:NR-KW-convolution}.
We use the $(++)$ and $(+-)$ channels as representatives for
$A_q$ and $D_q$, respectively.
Whereas $\mathcal{K}_A^{(++)}$ is regular, cancellation of
the separate $\mathcal O(v^{-2})$ terms in the $D_q$
subtraction leaves
\begin{equation}\label{eq:NR-Kpole-summary}
  \mathcal{K}_D^{(+-)}
  =
  -\frac{8m_Qk_z}{\bm\Delta_\perp^2}
  +\frac12+\mathcal O(v).
\end{equation}
The first term is odd in $k_z$ and of order $v^{-1}$,
which determines the required order of the overlap expansion.

At fixed scaled momenta, the kernels and overlap expand as
\begin{align}
  \mathcal{K}_A^{(++)}
  &=
  K_{A,0}+vK_{A,1}+v^2K_{A,2}+\cdots,
  \nonumber\\
  \mathcal{K}_D^{(+-)}
  &=
  v^{-1}K_{D,-1}+K_{D,0}
  +vK_{D,1}+v^2K_{D,2}+\cdots,
  \nonumber\\
  \mathcal W
  &=
  W_0+vW_1+v^2W_2+v^3W_3+\cdots.
  \label{eq:NR-KW-series}
\end{align}
Here $K_{F,n}$ denotes the expansion coefficients of
$\mathcal{K}_F$.
For each radial contribution $r=1S,2S,{\rm cross}$, we write
$W_n^{(r)}=\mathcal W_{G,0}\,w_n^{(r)}$ in the notation of
Sec.~\ref{Sec:NRQCD-W}.
The common Gaussian weight $\mathcal W_{G,0}$ carries the
dimensional factor $\beta^{-3}$, while $w_n^{(r)}$ is
dimensionless and includes the appropriate node factors.
Thus, $W_0^{(r)}$ denotes the leading overlap of the selected
radial contribution and is not necessarily a pure Gaussian.

Suppressing the channel and radial labels, and denoting the
coefficient of $v^n$ by $[\cdots]_{v^n}$, the required
products are
\begin{align}
  [\mathcal{K}_A\mathcal W]_{v^2}
  &=
  K_{A,2}W_0+K_{A,1}W_1+K_{A,0}W_2,
  \nonumber\\
  [\mathcal{K}_D\mathcal W]_{v^0}
  &=
  K_{D,0}W_0+K_{D,-1}W_1,
  \nonumber\\
  [\mathcal{K}_D\mathcal W]_{v^2}
  &=
  K_{D,2}W_0+K_{D,1}W_1
  +K_{D,0}W_2+K_{D,-1}W_3.
  \label{eq:NR-KW-products}
\end{align}
Since $W_0$ is even in $k_z$, the term
$v^{-1}K_{D,-1}W_0$ integrates to zero.
The product $K_{D,-1}W_1$, however, contributes at
$\mathcal O(v^0)$.
Consequently, $A_q$ through $\mathcal O(v^2)$ requires the
overlap through $W_2$, whereas the $(+-)$ extraction of
$D_q$ also requires $W_3$.
For the $2S$ and interference contributions, the node factors
must be expanded together with the Gaussian and Jacobian.
The expansion is therefore performed on the complete
kernel--overlap products with $x=x(k_z,\bm k_\perp)$,
rather than setting $x=1/2$ before integration.

%\subsection{Analytic form factors through \texorpdfstring{$\mathcal O(v^2)$}{order v2}}
%\label{Sec:NR-analytic-results}
Integrating the expanded kernel--overlap products gives the
diagonal and interference contributions to $A_q(t)$,
\begin{equation}\label{eq:NR-Aq-summary}
  \begin{aligned}
    A_q^{{\rm NR},\phi_{1S}}(t)
    &=
    \frac{G(\rho)}{2}
    \left[
      1+v^2\frac{\rho(\rho-24)}{256}
    \right]
    +\mathcal O(v^4),
    \\[1ex]
    A_q^{{\rm NR},\phi_{2S}}(t)
    &=
    \frac{G(\rho)}{2}
    \left[
      P_2(\rho)
      +v^2\frac{
      \rho(\rho^3-88\rho^2+2176\rho-13312)
      }{98304}
    \right]
    +\mathcal O(v^4),
    \\[1ex]
    A_q^{{\rm NR},{\rm cross}}(t)
    &=
    \frac{G(\rho)}{8\sqrt6}
    \left[
      \rho+v^2\frac{\rho^2(\rho-40)}{256}
    \right]
    +\mathcal O(v^4),
  \end{aligned}
\end{equation}
where $G(\rho)=e^{-\rho/16}=e^{t/(16\beta^2)}$ is the common
Gaussian factor from the leading overlap, and
$P_2(\rho)=1-\rho/12+\rho^2/384$ encodes the leading recoil
dependence of the $2S$ node product.
These expressions preserve the diagonal normalization
$A_q^{{\rm NR},\phi_{nS}}(0)=1/2$ and the vanishing interference
$A_q^{{\rm NR},{\rm cross}}(0)=0$.

For $D_q$, the two leading-order products in
Eq.~\eqref{eq:NR-KW-products} cancel after integration for
the pure Gaussian $1S$ and interference contributions,
but not for the diagonal $2S$ contribution.
Including all $\mathcal O(v^2)$ products gives
\begin{equation}\label{eq:NR-Dq-summary}
  \begin{aligned}
    D_q^{{\rm NR},\phi_{1S}}(t)
    &=
    -v^2G(\rho)+\mathcal O(v^4),
    \\[1ex]
    D_q^{{\rm NR},\phi_{2S}}(t)
    &=
    -\frac23G(\rho)
    \left[
      1+\frac{v^2}{128}
      \left(\rho^2-44\rho+832\right)
    \right]
    +\mathcal O(v^4),
    \\[1ex]
    D_q^{{\rm NR},{\rm cross}}(t)
    &=
    -\frac4{\sqrt6}\,v^2G(\rho)
    \left(1+\frac{\rho}{16}\right)
    +\mathcal O(v^4).
  \end{aligned}
\end{equation}
At this order, the pure $1S$ contribution follows the Gaussian
recoil dependence, whereas the $2S$ correction and interference
contain additional polynomial dependence on $\rho$.

The interference admits an independent check through the
common-width HO identity derived in
Sec.~\ref{Sec:NRQCD-W}.
At fixed $m_Q$, $t$, and $\vec k$, the extraction kernels,
recoil relations, and Jacobian have no explicit $\beta$
dependence.
Differentiating the pure $1S$ convolution therefore gives,
for either $F=A$ or $D$,
\begin{equation}\label{eq:NR-HO-width-cross}
  F_q^{\rm cross}(t)
  =
  \frac2{\sqrt6}\,\beta
  \left.
    \frac{\partial F_q^{\phi_{1S}}(t)}{\partial\beta}
  \right|_{m_Q,t}.
\end{equation}
This identity holds for the full fBT convolution and reproduces
both interference expressions above order by order.
When applied to the NR expressions, the derivative acts on
both $v=\beta/m_Q$ and $\rho=-t/\beta^2$.
It is taken at fixed $m_Q$ and $t$, not at fixed $\rho$.

The exact kernel identity in Eq.~\eqref{eq:A-universal}
ensures current-component independence of the $A_q$ expansion.
For $D_q$, the transverse-trace kernel
$\mathcal K_D^{(ii)}$ has no $v^{-1}$ term, unlike the common
$(+-)$ and contracted $(-i)$ kernels.
Its convolution, including the $e^i$ contribution of
Table~\ref{tab:trace}, nevertheless reproduces all three
results in Eq.~\eqref{eq:NR-Dq-summary} through
$\mathcal O(v^2)$, as verified in Sec.~\ref{Sec:NRQCD-D}.

%\subsection{Physical mixtures and forward limits}
%\label{Sec:NR-physical-mixtures}

For either $F=A$ or $D$, bilinearity in the LFWFs gives the
form factors of the physical mixtures in Eq.~\eqref{eq:HOWF1},
\begin{subequations}\label{eq:NR-Fq-mix-summary}
\label{eq:NR-Aq-mix}
  \begin{align}
    F_q^{\Phi_{1S}}(t)
    &=
    c_1^2 F_q^{\phi_{1S}}(t)
    +c_2^2 F_q^{\phi_{2S}}(t)
    +c_1c_2 F_q^{\rm cross}(t),
    \label{eq:NR-Fq-mix-1S-summary}
    \\
    F_q^{\Phi_{2S}}(t)
    &=
    c_2^2 F_q^{\phi_{1S}}(t)
    +c_1^2 F_q^{\phi_{2S}}(t)
    -c_1c_2 F_q^{\rm cross}(t),
    \label{eq:NR-Fq-mix-2S-summary}
  \end{align}
\end{subequations}
where $c_1=\cos\theta$ and $c_2=\sin\theta$, with
$c_1^2+c_2^2=1$.
The interference form factor $F_q^{\rm cross}$ includes both
ordered overlaps, so no additional factor of two is required.
Substituting Eqs.~\eqref{eq:NR-Aq-summary}
and~\eqref{eq:NR-Dq-summary} gives the mixed-state form factors
through $\mathcal O(v^2)$.
The corresponding total valence form factors are
$F^{\Phi_{nS}}(t)=2F_q^{\Phi_{nS}}(t)$.

The orthogonal mixing preserves $A_q^{\Phi_{nS}}(0)=1/2$
and hence $A^{\Phi_{nS}}(0)=1$.
The forward $D_q$ values through $\mathcal O(v^2)$ are
\begin{equation}\label{eq:NR-D-mix}
  \begin{aligned}
    D_q^{{\rm NR},\Phi_{1S}}(0)
    &=
    -\frac23c_2^2
    -v^2\left[
      c_1^2+\frac{13}{3}c_2^2
      +\frac4{\sqrt6}c_1c_2
    \right]
    +\mathcal O(v^4),
    \\[1ex]
    D_q^{{\rm NR},\Phi_{2S}}(0)
    &=
    -\frac23c_1^2
    -v^2\left[
      c_2^2+\frac{13}{3}c_1^2
      -\frac4{\sqrt6}c_1c_2
    \right]
    +\mathcal O(v^4).
  \end{aligned}
\end{equation}
At fixed mixing coefficients, the leading forward terms are
independent of $m_Q$ and $\beta$.
The $\mathcal O(v^2)$ corrections introduce scale dependence
through $v^2=\beta^2/m_Q^2$ and include both diagonal and
interference contributions.
The interference enters the two physical states with opposite signs.

For a nonzero $c_2$ held fixed as $v\to0$, the physical ground
state therefore retains an $\mathcal O(v^0)$ $D$-term,
whereas the pure Gaussian $1S$ contribution begins at
$\mathcal O(v^2)$.
This Gaussian suppression does not follow from nodelessness alone.
Section~\ref{Sec:NRQCD-D} gives a nodeless momentum-space
mixture with a nonzero leading $D_q(0)$.
The cancellation for the pure Gaussian is a bound-state result
within the present constituent-current prescription and should
not be identified with the free-Dirac result
$D_{\rm Dirac}(t)=0$~\cite{Hudson:2017oul}.

\section{Numerical Results}\label{Sec:V}

\begin{table*}[t]
\centering
\caption{Wave-function parameters of Ref.~\cite{ACJO22}
used for the physical $1S$ and $2S$ mixtures with $\theta=12^\circ$.
The masses, the potential constant $a$, and the Gaussian widths are
in GeV, while $\alpha_s$ is dimensionless.}
\renewcommand{\arraystretch}{1.25}
\setlength{\tabcolsep}{9pt}
\begin{tabular}{llcccccc}
\hline\hline
Model & Mixing scenario & $m_c$ & $m_b$ & $a$ & $\alpha_s$ & $\beta_{cc}$ & $\beta_{bb}$ \\
\hline
Model 1~\cite{ACJO22} & Mixed ($\theta=12^\circ$) & 1.68 & 5.10 & $-0.543$ & 0.433 & 0.592 & 1.167 \\
\hline\hline
\end{tabular}
\label{table:wf_parameters}
\end{table*}

We present the total valence form factors $A(t)=2A_q(t)$ and
$D(t)=2D_q(t)$ for the physical $\eta_c(1S,2S)$ and $\eta_b(1S,2S)$
states, obtained from Eqs.~\eqref{eq:Aextract} and~\eqref{eq:Dextract}.
All comparisons use the same parameters in
Table~\ref{table:wf_parameters} and the same orthogonal mixing
with $\theta=12^\circ$.
Thus, the $1S$ and $2S$ labels in the tables and figures denote
physical mixtures, not the unmixed HO basis states.
For these parameters,
\[
  v_c=\frac{\beta_{cc}}{m_c}\simeq0.352,
  \qquad v_c^2\simeq0.124,
  \qquad
  v_b=\frac{\beta_{bb}}{m_b}\simeq0.229,
  \qquad v_b^2\simeq0.0524.
\]
Bottomonium therefore has a smaller NR expansion parameter even
though its absolute momentum width is larger.
The width ratio $\beta_{bb}/\beta_{cc}\simeq1.97$ and the velocity
ratio describe different aspects of the calculation.
The former sets the characteristic recoil and spatial scales,
whereas the latter controls the velocity expansion.

\subsection{Gravitational form factors}

\begin{figure}[!htbp]
    \centering
    \includegraphics[width=0.48\linewidth]{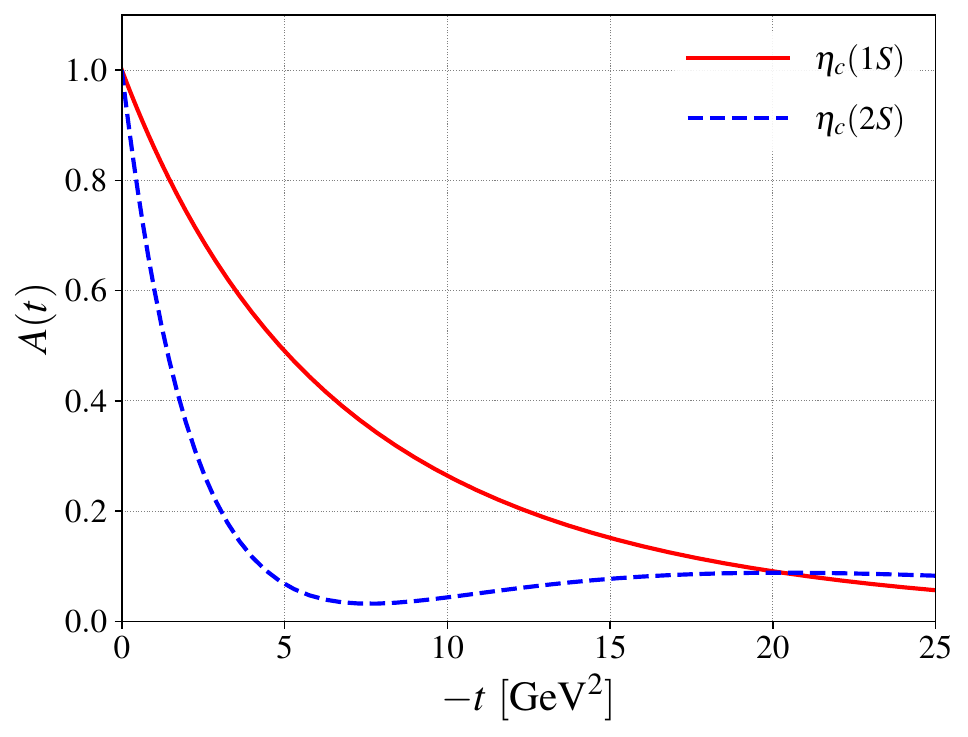}
    \includegraphics[width=0.48\linewidth]{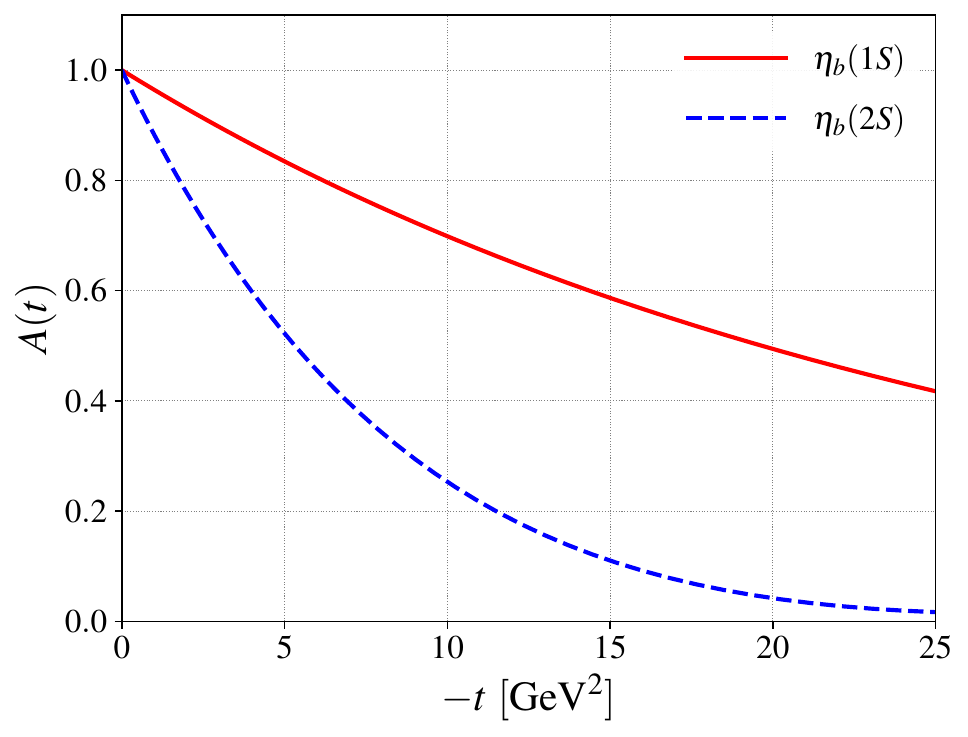}
    \caption{Full relativistic total valence form factor
    $A(t)$ for $\eta_c$ (left) and $\eta_b$ (right).
    Red solid and blue dashed curves denote the physical $1S$
    and $2S$ mixtures, respectively, with $\theta=12^\circ$ and
    the parameters of Table~\ref{table:wf_parameters}.}
    \label{fig:GFF_A}
\end{figure}

Figure~\ref{fig:GFF_A} shows the flavor and radial-state dependence
of the full relativistic form factor $A(t)$.
Near the forward limit, both bottomonium states exhibit a slower
decrease with increasing $-t$ than their charmonium counterparts.
At a fixed momentum transfer, however, both $v$ and
$\rho=-t/\beta^2$ differ between the two flavors, so this comparison
does not isolate the constituent-mass dependence alone.
Within each flavor, the $2S$ curve initially decreases more steeply
than the $1S$ curve.
Since the two states share the same $\beta$, their different
profiles reflect the radial-node and interference structure
rather than a separately adjusted width.

The $\eta_c(2S)$ curve develops a positive minimum in the region
$-t\simeq5$--$10$~GeV$^2$ and then rises into a broad shoulder.
This nonmonotonic behavior is consistent with cancellations between
recoil-shifted regions on opposite sides of the radial node,
together with the $1S$--$2S$ interference.
A similar sensitivity to radial structure was found for the vector
form factors in Ref.~\cite{ACJO22}.
No corresponding minimum is resolved for $\eta_b(2S)$ in the
plotted interval.
The same value of $\rho$ corresponds to a momentum transfer about
$\beta_{bb}^2/\beta_{cc}^2\simeq3.89$ times larger in bottomonium,
although the simultaneous change in $v$ prevents an exact rescaling
of the two full curves.
The large-recoil structure in Fig.~\ref{fig:GFF_A} follows from
the full convolution and is not inferred by extrapolating the
low-order NR expansion to that region.

\begin{figure}[!htbp]
    \centering
    \includegraphics[width=0.48\linewidth]{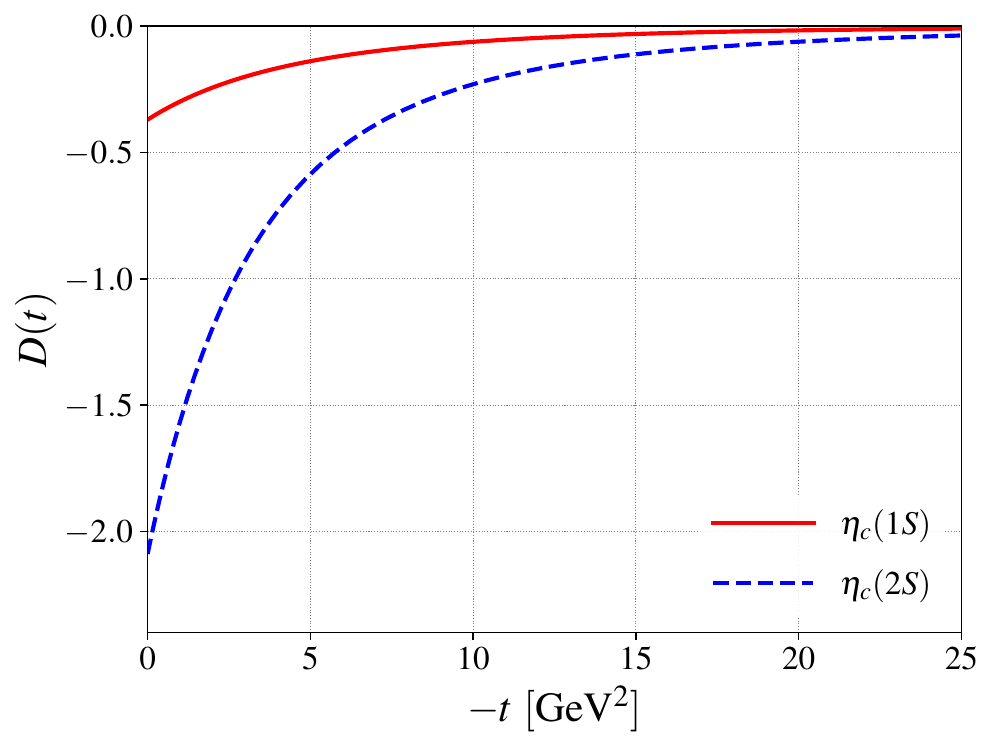}
    \includegraphics[width=0.48\linewidth]{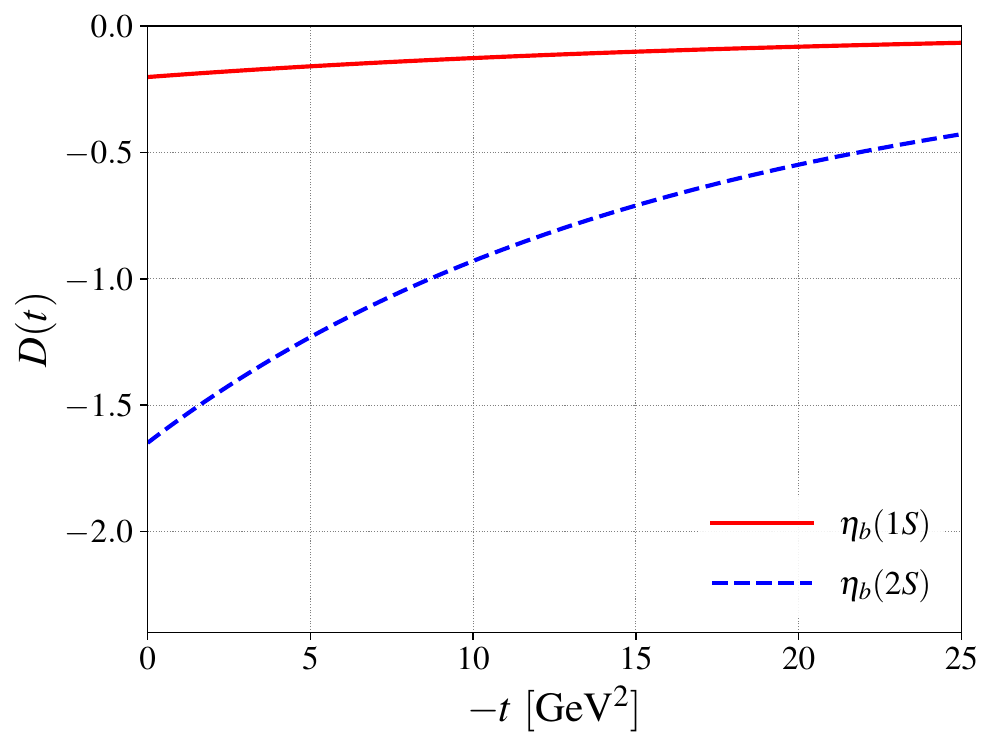}
    \caption{Full relativistic total valence form factor
    $D(t)$ for $\eta_c$ (left) and $\eta_b$ (right).
    The state and parameter conventions are the same as in
    Fig.~\ref{fig:GFF_A}.}
    \label{fig:GFF_D}
\end{figure}

Figure~\ref{fig:GFF_D} shows the corresponding $D$-term form factors.
All four curves remain negative over the displayed spacelike
interval and decrease in magnitude as $-t$ increases.
For both flavors, the $2S$ state has a substantially larger
$|D(0)|$ than the $1S$ state.
The charmonium form factors approach zero at relatively small
values of $-t$, whereas the bottomonium form factors retain
appreciable magnitudes up to larger momentum transfers.
Unlike $A(t)$ for $\eta_c(2S)$, the displayed $D(t)$ curves show
no analogous minimum-and-shoulder structure.

\begin{table*}[t]
\centering
\caption{Forward values and slopes of the total valence
GFFs for the physical mixtures with $\theta=12^\circ$.
All $A(0)=1$, the $D(0)$ entries are dimensionless, and all slopes
are in GeV$^{-2}$.
The subscript $\mathrm{NR}$ denotes the analytic approximation
through $\mathcal O(v^2)$, evaluated from
Eqs.~\eqref{eq:NR-Aq-summary}, \eqref{eq:NR-Dq-summary},
and~\eqref{eq:NR-Fq-mix-summary}.
The additional NR $D$ values and slopes allow the normalization
and recoil dependence to be compared separately.}
\renewcommand{\arraystretch}{1.25}
\setlength{\tabcolsep}{6pt}
\begin{tabular}{llcccccc}
\hline\hline
Meson & State & $A'(0)$ & $A_{\mathrm{NR}}'(0)$ & $D(0)$ &
{$D_{\mathrm{NR}}(0)$} & $D'(0)$ &
{$D_{\mathrm{NR}}'(0)$} \\
\hline
$\eta_c$ & $1S$ & 0.156 & {0.163} & $-0.370$ & {$-0.424$} & $-0.082$ & {$-0.068$} \\
  & $2S$ & {0.501} & {0.512} & $-2.088$ & {$-2.234$} & $-0.627$ & {$-0.568$} \\
\hline
$\eta_b$ & $1S$ & 0.037 & {0.037} & $-0.201$ & {$-0.212$} & $-0.010$ & {$-0.009$} \\
  & $2S$ & 0.124 & {0.125} & $-1.649$ & {$-1.680$} & $-0.099$ & {$-0.096$} \\
\hline\hline
\end{tabular}
\label{table:gff_summary}
\end{table*}

%%%%%
Table~\ref{table:gff_summary} quantifies the forward behavior
of the two form factors.
The full relativistic values and slopes are compared with the
analytic NR approximation through $\mathcal O(v^2)$, using
the same physical mixtures and parameter sets.

For the full calculation, the $2S$ slopes $A'(0)$ are
approximately $3.2$ and $3.4$ times the corresponding $1S$
values for charmonium and bottomonium, respectively.
Since all states satisfy $A(0)=1$, these ratios directly
quantify the stronger initial decrease of the excited-state
curves in Fig.~\ref{fig:GFF_A}.

The full relativistic calculation gives the forward $D$-terms
\[
  \begin{aligned}
    D_{\eta_c(1S)}(0)&=-0.370,
    &\qquad D_{\eta_c(2S)}(0)&=-2.088,
    \\
    D_{\eta_b(1S)}(0)&=-0.201,
    &\qquad D_{\eta_b(2S)}(0)&=-1.649.
  \end{aligned}
\]
These are the total valence form factors, including both
the quark and antiquark contributions.
For a given radial state, the bottomonium value is smaller
in magnitude than the charmonium value.
Radial excitation produces a much larger change,
with $|D_{2S}(0)|/|D_{1S}(0)|\simeq5.64$ for charmonium
and $8.20$ for bottomonium.

For $\eta_c(1S)$, our value can be compared with the BLFQ
results $D(0)\simeq-4.5$ and $-4.7$ obtained from
$T^{12}$ and $T^{+-}$, respectively~\cite{Xu:2024hfx,Hu:2024edc}.
Our ground-state magnitude is therefore more than an order
of magnitude smaller.
For $\eta_c(2S)$, the corresponding BLFQ central values are
approximately $-6.5$ and $-6.7$~\cite{Hu:2024edc},
compared with our result $-2.088$.
Thus, both calculations find an increase in $|D(0)|$
under radial excitation, but the enhancement relative to
the ground state is considerably stronger in the present
calculation.
In BLFQ the $2S/1S$ magnitude ratio is about $1.4$,
whereas our charmonium ratio is about $5.6$.

The contact-interaction calculation based on coupled
Dyson--Schwinger and Bethe--Salpeter equations gives
$D_{\eta_c(1S)}(0)=-0.576(3)$ and
$D_{\eta_b(1S)}(0)=-0.445(13)$~\cite{Sultan:2024hep}.
Our ground-state magnitudes are approximately $36\%$ and
$55\%$ smaller than these central values, respectively.
The two approaches therefore share the trend of decreasing
$|D(0)|$ from charmonium to bottomonium, but differ
quantitatively.
In particular, our $\eta_c(1S)$ result is numerically closer
to the contact-interaction prediction than to the BLFQ
values, without implying quantitative agreement.

The normalized slopes provide complementary information
about the momentum-transfer dependence near $t=0$.
The ratios $D'(0)/D(0)$ are approximately
$0.222$ and $0.300$~GeV$^{-2}$ for
$\eta_c(1S)$ and $\eta_c(2S)$, compared with
$0.050$ and $0.060$~GeV$^{-2}$ for
$\eta_b(1S)$ and $\eta_b(2S)$.
Their increase under radial excitation is much smaller
than the increase in $|D(0)|$.
The excited states therefore have substantially larger
forward magnitudes, while their fractional decrease
near $t=0$ is only moderately faster.
The forward normalization and the normalized slope
thus distinguish the overall magnitude from the initial
relative momentum-transfer dependence.

To assess the NR approximation to these forward quantities,
we first consider $A'(0)$.
The full and NR results all satisfy $A(0)=1$, so agreement
at the origin alone does not establish the accuracy of
the momentum-transfer dependence.
For the diagonal HO contributions and their interference,
we use the total valence NR form factors
$F^{{\rm NR},r}(t)=2F_q^{{\rm NR},r}(t)$,
with $F=A,D$ and $r=\phi_{1S},\phi_{2S},{\rm cross}$.
Differentiating the retained terms of
Eq.~\eqref{eq:NR-Aq-summary} gives the slopes through
$\mathcal O(v^2)$,
\begin{equation}\label{eq:NR-A-slopes}
  \begin{aligned}
    a_1
    &\equiv
    \left.
      \frac{dA^{{\rm NR},\phi_{1S}}(t)}{dt}
    \right|_{t=0}
    =
    \frac{1}{16\beta^2}+\frac{3}{32m_Q^2},
    \\[1ex]
    a_2
    &\equiv
    \left.
      \frac{dA^{{\rm NR},\phi_{2S}}(t)}{dt}
    \right|_{t=0}
    =
    \frac{7}{48\beta^2}+\frac{13}{96m_Q^2},
    \\[1ex]
    a_{\rm cross}
    &\equiv
    \left.
      \frac{dA^{{\rm NR},{\rm cross}}(t)}{dt}
    \right|_{t=0}
    =
    -\frac{1}{4\sqrt6\,\beta^2}.
  \end{aligned}
\end{equation}
Here $a_1$ and $a_2$ refer to the diagonal HO basis
contributions, while $a_{\rm cross}$ is the slope of the
sum of the two ordered interference contributions.
All slopes are defined with respect to $t$ and have
units of GeV$^{-2}$.

The slopes of the physical mixtures,
$a_{\Phi_n}\equiv
[dA^{{\rm NR},\Phi_{nS}}(t)/dt]_{t=0}$, are
\[
  \begin{aligned}
    a_{\Phi_1}
    &=c_1^2a_1+c_2^2a_2+c_1c_2a_{\rm cross},
    \\
    a_{\Phi_2}
    &=c_2^2a_1+c_1^2a_2-c_1c_2a_{\rm cross}.
  \end{aligned}
\]
These give the entries $A'_{\mathrm{NR}}(0)$ in
Table~\ref{table:gff_summary}.
Since the adopted mixing convention has $c_1c_2>0$,
the negative interference slope lowers the ground-state
slope and raises the excited-state slope relative to
the diagonal contributions alone.
Although $c_2^2\simeq0.0432$, the interference coefficient
is $c_1c_2\simeq0.2034$, so the effect of mixing cannot
be described by the small admixture probability alone.

For $\eta_c(1S)$ and $\eta_c(2S)$,
Table~\ref{table:gff_summary} gives the NLO slopes
$A'_{\mathrm{NR}}(0)=0.163$ and $0.512$~GeV$^{-2}$,
compared with the full-calculation values
$A'(0)=0.156$ and $0.501$~GeV$^{-2}$, respectively.
The NLO slopes therefore exceed the full results by
approximately $4.5\%$ and $2.2\%$.
For $\eta_b(1S)$ and $\eta_b(2S)$, the tabulated NLO
slopes are $0.037$ and $0.125$~GeV$^{-2}$,
compared with the full values $0.037$ and
$0.124$~GeV$^{-2}$, respectively.
The bottomonium slopes thus show closer agreement.
The two ground-state entries coincide at the displayed
precision, which should not be interpreted as exact equality.
These comparisons use the rounded values in
Table~\ref{table:gff_summary}.

For $D(t)$, the forward value and the slope provide separate
tests because the normalization is not fixed.
With the common mixing angle, Eq.~\eqref{eq:NR-D-mix}
gives the total leading values
$D_{\rm LO}^{\Phi_{1S}}(0)\simeq-0.0576$ and
$D_{\rm LO}^{\Phi_{2S}}(0)\simeq-1.2757$,
independently of flavor.
These leading terms are proportional to $c_2^2$ and $c_1^2$,
respectively, explaining the strong radial-state dependence
within the present prescription.
The respective $\mathcal O(v^2)$ corrections are approximately
$-2.9524v^2$ and $-7.7143v^2$.
The ground-state correction is larger than its small
LO contribution at both physical parameter sets.
This reflects the Gaussian LO cancellation and the mixing-induced
suppression of the surviving leading term, rather than an
inconsistency of the common power counting.
The excited state has a larger absolute NLO coefficient,
but its unsuppressed LO term makes the relative correction smaller.

Numerically, the ground-state NLO correction is about $-0.367$
for charmonium and $-0.155$ for bottomonium.
The interference contributes about $-0.0825$ and $-0.0348$
to these shifts, respectively, and has the opposite sign for
the physical $2S$ states.
The NLO forward values in Table~\ref{table:gff_summary}
overestimate $|D(0)|$ by approximately $15\%$ and $7\%$
for $\eta_c(1S,2S)$, compared with about $6\%$ and $2\%$
for $\eta_b(1S,2S)$.
Thus, the small deviations in $A'_{\mathrm{NR}}(0)$ do not
imply the same accuracy for $D_{\mathrm{NR}}(0)$.
The stronger relative flavor dependence of the ground-state
$D$-term also explains why $|D_{2S}(0)|/|D_{1S}(0)|$ is larger
for bottomonium despite its smaller overall relativistic corrections.

The $D$-term slopes follow by differentiating the retained terms
of Eq.~\eqref{eq:NR-Dq-summary},
\begin{equation}\label{eq:NR-D-slopes}
  \begin{aligned}
    d_1
    &\equiv
    \left.
      \frac{dD^{{\rm NR},\phi_{1S}}(t)}{dt}
    \right|_{t=0}
    =
    -\frac{1}{8m_Q^2},
    \\[1ex]
    d_2
    &\equiv
    \left.
      \frac{dD^{{\rm NR},\phi_{2S}}(t)}{dt}
    \right|_{t=0}
    =
    -\frac{1}{12\beta^2}-\frac{1}{m_Q^2},
    \\[1ex]
    d_{\rm cross}
    &\equiv
    \left.
      \frac{dD^{{\rm NR},{\rm cross}}(t)}{dt}
    \right|_{t=0}
    =0.
  \end{aligned}
\end{equation}
These expressions use the same total-valence convention and
NLO truncation as Eq.~\eqref{eq:NR-A-slopes}.
The vanishing interference slope follows from the cancellation
between the derivatives of $G(\rho)$ and $1+\rho/16$ at $\rho=0$,
not from a vanishing interference form factor.
Defining
$d_{\Phi_n}\equiv[dD^{{\rm NR},\Phi_{nS}}(t)/dt]_{t=0}$,
the physical slopes reduce at this order to
\[
  d_{\Phi_1}=c_1^2d_1+c_2^2d_2,
  \qquad
  d_{\Phi_2}=c_2^2d_1+c_1^2d_2.
\]
These give the entries $D'_{\mathrm{NR}}(0)$ in
Table~\ref{table:gff_summary}.
Although the interference slope vanishes through $\mathcal O(v^2)$,
its contribution to $D_{\mathrm{NR}}(0)$ still affects the
normalized slope $D'_{\mathrm{NR}}(0)/D_{\mathrm{NR}}(0)$.

The NLO slopes in Table~\ref{table:gff_summary} underestimate
the magnitudes of the full $D'(0)$ values, whereas the NLO
magnitudes of $D(0)$ are overestimated.
Both deviations lower $D'_{\mathrm{NR}}(0)/D_{\mathrm{NR}}(0)$
relative to the full normalized slope.
Thus, the accuracy of the NR description must be assessed
separately for $A'(0)$, $D(0)$, and $D'(0)$.
The ratios and percentage comparisons above use the displayed
precision of Table~\ref{table:gff_summary} and the central
values of the cited calculations.

%%%%% Fig. 3
\begin{figure}[!htbp]
    \centering
    \includegraphics[width=0.9\linewidth]{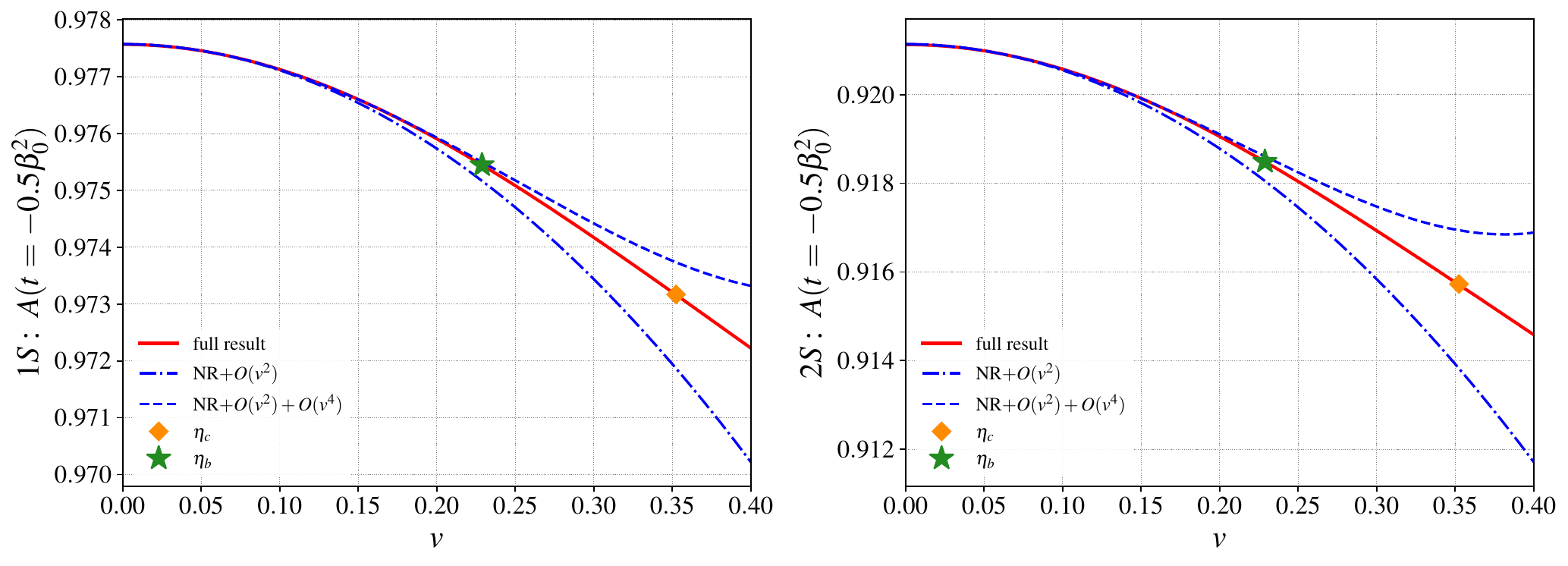}
    \includegraphics[width=0.9\linewidth]{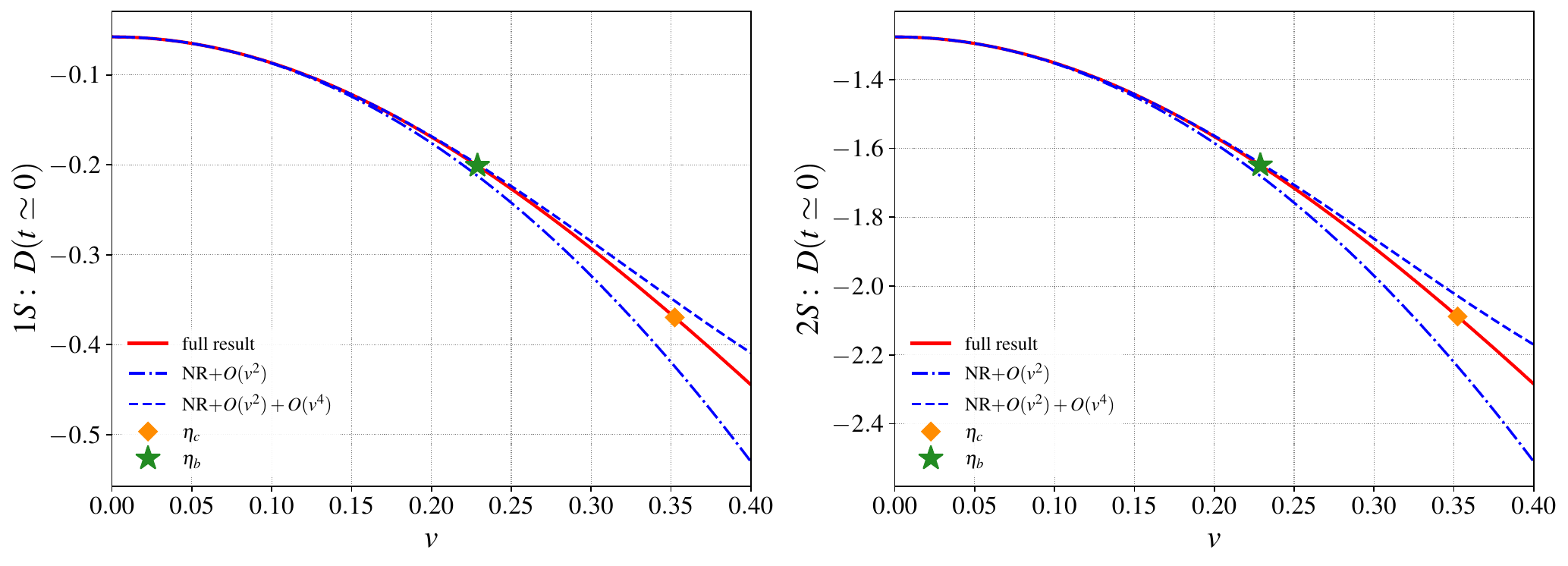}
    \caption{Velocity dependence of the total valence GFFs
    for the physical $1S$ (left) and $2S$ (right) mixtures
    with $\theta=12^\circ$.
    The upper row shows $A(t)$ at $\rho=0.5$, and the lower
    row shows $D(t)$ near the forward limit.
    Red solid, blue dash-dotted, and blue dashed curves
    denote the full calculation, the analytic NLO approximation
    through $\mathcal O(v^2)$, and the NLO approximation
    supplemented by the numerically evaluated
    $\mathcal O(v^4)$ correction, respectively.
    For the upper row, all curves are evaluated with
    $\beta=\beta_0=1$~GeV at fixed $\rho=0.5$.
    The orange diamond and green star mark the physical
    charmonium and bottomonium results at $v_c\simeq0.352$
    and $v_b\simeq0.229$, respectively.
    In the upper row, these points use their physical widths
    and correspond to $t=-0.5\beta_{cc}^2$ and
    $t=-0.5\beta_{bb}^2$.}
    \label{fig:GFFs_vs_NR}
\end{figure}
%%%%%%%%%%%%%%%%%%%
Figure~\ref{fig:GFFs_vs_NR} compares the velocity dependence
of the full relativistic results with the NR approximations
at fixed $\rho=0.5$ for $A(t)$ and near the forward limit
for $D(t)$.
The red solid curves show the full relativistic results,
while the blue curves represent the NR approximations.

For $A(t)$ in the upper panels, both the full and NR curves
are evaluated with $\beta=\beta_0=1$~GeV and
$t=-0.5\beta_0^2$.
The physical charmonium and bottomonium results, marked by
the orange diamond and green star, respectively, are calculated
using their own parameter sets.
To maintain the same $\rho=0.5$, these points are evaluated
at $t=-0.5\beta_{cc}^2\simeq-0.175$~GeV$^2$ and
$t=-0.5\beta_{bb}^2\simeq-0.681$~GeV$^2$, respectively.

For each radial state in the present equal-mass HO model,
rescaling the integration momenta by $\beta$ leaves the
dimensionless form factors dependent only on $v$ and $\rho$
at fixed $\theta$.
The physical points in the upper panels therefore lie on
the corresponding full relativistic curves at their
respective values of $v=\beta/m_Q$, despite the different
widths and physical momentum transfers.

In the lower panels, each radial state approaches its own
mixing-dependent LO value as $v\to0$.
These limiting values are independent of flavor and are
given by Eq.~\eqref{eq:NR-D-mix}.
The deviations from the leading values become larger as
$v$ increases toward the charmonium point.
This is consistent with the stronger relativistic corrections
to the forward $D$-terms discussed with
Table~\ref{table:gff_summary}.

Including the numerically evaluated $\mathcal O(v^4)$
correction brings the approximations closer to the full
results around both physical points.
The NLO and fourth-order approximations can lie on opposite
sides of the full result.
Their separation is more visible near the charmonium point,
particularly in the lower panels, than near the bottomonium
point.
The factors $v_c^4\simeq0.0154$ and $v_b^4\simeq0.00274$
show the different nominal suppression of the fourth-order
contributions.
Their actual sizes also depend on the radial-state and
momentum-transfer-dependent coefficients.
The observed improvement concerns the velocity range and
momentum-transfer conditions shown here, rather than the
large-$-t$ behavior in Figs.~\ref{fig:GFF_A}
and~\ref{fig:GFF_D}.

The improvement is weakest for $A(t)$ of the $2S$ state at the
charmonium point, where the NLO and $\mathcal O(v^4)$ approximations
lie on opposite sides of the full result at comparable distances,
the relative deviation decreasing only from $0.21\%$ to $0.13\%$.
{For the pure HO basis states, the ratio of the magnitudes
of the $\mathcal O(v^4)$ and $\mathcal O(v^2)$ coefficients
of the forward slope of $A(t)$ is
$93/26\simeq3.6$ for $\phi_{2S}$, compared with $11/6\simeq1.8$ for
$\phi_{1S}$.}
This reflects the larger high-momentum content of the nodal wave
function,
$\langle\vec k^{\,4}\rangle_{\phi_{2S}}/\langle\vec k^{\,4}\rangle_{\phi_{1S}}=5$
compared with $7/3$ for $\langle\vec k^{\,2}\rangle$.
The velocity expansion therefore converges slowly for $\eta_c(2S)$
at $v_c^2\simeq0.124$, whereas for $\eta_b(2S)$, with
$v_b^2\simeq0.0524$, the $\mathcal O(v^4)$ term improves the
approximation as in the other cases.

Table~\ref{table:gff_radii} presents the length scales
associated with the form factors.
The mass and $D$-term slope radii follow from the forward
quantities in Table~\ref{table:gff_summary}, while the
mechanical radius is evaluated using the full spacelike
form factor $D(t)$.

\begin{table*}[t]
\centering
\caption{Radii of the physical $S$-wave heavy quarkonia
in fm.
The $A$ and $D$ slope radii use
$\langle r_F^2\rangle=6F'(0)/F(0)$, while the mechanical
radius uses Eq.~\eqref{eq:rmech}.
The vector radii are quoted from Ref.~\cite{ACJO22}.
The gravitational radii are obtained from the full
relativistic calculation, except for the NR mass radii,
which use the analytic NLO slopes of the same physical mixtures
in Table~\ref{table:gff_summary}.}
\renewcommand{\arraystretch}{1.25}
\setlength{\tabcolsep}{9pt}
\begin{tabular}{llccccc}
\hline\hline
Meson & State
& $\sqrt{\langle r_V^2\rangle}$\cite{ACJO22}
& $\sqrt{\langle r_A^2\rangle}$
& $\sqrt{\langle r_A^2\rangle_{\mathrm{NR}}}$
& $\sqrt{\langle r_D^2\rangle}$
& $\sqrt{\langle r^2\rangle_{\mathrm{mech}}}$ \\
\hline
$\eta_c$ & $1S$
& 0.205 & 0.191 & {0.195} & 0.228 & 0.202 \\
& $2S$
& 0.359 & {0.342} & {0.346} & 0.265 & 0.227 \\
\hline
$\eta_b$ & $1S$
& 0.095 & 0.092 & {0.093} & 0.106 & 0.100 \\
& $2S$
& 0.173 & 0.170 & {0.171} & 0.118 & 0.109 \\
\hline\hline
\end{tabular}
\label{table:gff_radii}
\end{table*}

The NR mass radii reflect the slope comparison in
Table~\ref{table:gff_summary}.
For $\eta_c(1S)$ and $\eta_c(2S)$, the NLO radii
$0.195$ and $0.346$~fm exceed the full values
$0.191$ and $0.342$~fm by approximately $2\%$ and
$1\%$, respectively.
% For bottomonium, the ground-state values coincide at
% the displayed precision, while the excited-state values
% are $0.171$~fm at NLO and $0.170$~fm in the full calculation.
For bottomonium, the NLO radii $0.093$ and $0.171$~fm 
exceed the full-calculation values $0.092$ and $0.170$~fm 
by less than 1$\%$.
Since the mass radius is proportional to $\sqrt{A'(0)}$,
its fractional deviation is approximately half that of
the slope when the deviation is small.
This agreement concerns the mass radius and does not
imply the same accuracy for quantities derived from $D(t)$.

The full-calculation radii in Table~\ref{table:gff_radii}
show that radial excitation affects the three gravitational
length scales differently.
For charmonium, the mass radius increases from $0.191$~fm
for the physical $1S$ state to $0.342$~fm for $2S$.
In comparison, the $D$-term slope radius increases from
$0.228$ to $0.265$~fm, while the mechanical radius changes
from $0.202$ to $0.227$~fm.
The same pattern appears in bottomonium.
The mass radius increases from $0.092$~fm for $1S$ to
$0.170$~fm for $2S$, whereas the $D$-term slope radius
increases from $0.106$ to $0.118$~fm and the mechanical
radius from $0.100$ to $0.109$~fm.
For both flavors, the mass radius is therefore substantially
more sensitive to radial excitation than either the
$D$-term slope radius or the mechanical radius.

Comparing the two flavors, the charmonium mass radii are
approximately twice the corresponding bottomonium values
for both radial states.
This pattern is consistent with the larger Gaussian width
of bottomonium, $\beta_{bb}=1.167$~GeV compared with
$\beta_{cc}=0.592$~GeV in Table~\ref{table:wf_parameters},
which corresponds to a smaller characteristic spatial extent.
At fixed radial state and mixing angle, the leading NR
mass radius scales as $1/\beta$.
The full relativistic result also depends on $v=\beta/m_Q$,
so the different values of $v$ for the two flavors modify
this simple inverse-width relation.

For both flavors, the radii obtained from the full
relativistic calculation satisfy
\begin{align}
  &\sqrt{\langle r_A^2\rangle}
   <\sqrt{\langle r^2\rangle_{\mathrm{mech}}}
   <\sqrt{\langle r_D^2\rangle}
   \qquad \text{for }1S,
  \\
  &\sqrt{\langle r_A^2\rangle}
   >\sqrt{\langle r_D^2\rangle}
   >\sqrt{\langle r^2\rangle_{\mathrm{mech}}}
   \qquad \text{for }2S.
\end{align}
The mass radius is therefore the smallest of the three
for the ground state and the largest for the excited state,
reflecting its stronger increase under radial excitation.
The mechanical radius remains smaller than the $D$-term
slope radius in both states.
Although both are obtained from $D(t)$, the mechanical
radius depends on the ratio of $D(0)$ to the integral of
$D(t)$ over the full spacelike region, whereas the slope
radius depends on $D'(0)/D(0)$.
They characterize different aspects of the momentum-transfer
dependence and need not coincide.

The ordering
$\sqrt{\langle r_A^2\rangle}<\sqrt{\langle r_D^2\rangle}$
for the $1S$ states is also found in the pion fBT-LFQM
calculation, which gives $0.546$ and $0.760$~fm,
respectively~\cite{Choi:2025}.
The vector radii in Table~\ref{table:gff_radii} exceed
all three gravitational radii for the $2S$ states.
For $1S$, the vector radius lies between the mechanical
and $D$-term slope radii in charmonium, but between
the mass and mechanical radii in bottomonium.
The BLFQ mass radii of approximately $0.18$ and
$0.35$~fm for $\eta_c(1S)$ and $\eta_c(2S)$,
respectively~\cite{Xu:2024hfx,Hu:2024edc},
are numerically close to the present values of
$0.191$ and $0.342$~fm.
The comparison uses the same three-dimensional
slope-radius convention as Eq.~\eqref{eq:radius}.
This similarity contrasts with the substantially different
$D(0)$ values discussed in connection with
Table~\ref{table:gff_summary}.

{Numerical evaluation confirms agreement among the three
current-component extractions for all four states.}

\subsection{Densities}\label{Sec:densities}

The transverse densities complement the form factors by resolving
the radial profiles associated with momentum and stress.
We use the two-dimensional definitions
\begin{align}\label{eq:2Ddensities}
    \mathcal E(x_\perp)
    &=P^+\widetilde A(x_\perp),
    \\
    \mathcal P(x_\perp)
    &=\frac{1}{4P^+}\frac{1}{2x_\perp}
      \frac{d}{dx_\perp}
      \left(
  x_\perp\frac{d}{dx_\perp}\widetilde D(x_\perp)
      \right),
    \\
    \mathcal S(x_\perp)
    &=-\frac{1}{4P^+}x_\perp
      \frac{d}{dx_\perp}
      \left(
     \frac{1}{x_\perp}\frac{d}{dx_\perp}
     \widetilde D(x_\perp)
    \right),
\end{align}
where the two-dimensional Fourier transform is
\begin{equation}
    \widetilde F(x_\perp)
    =
    \int\frac{d^2\bm\Delta_\perp}{(2\pi)^2}\,
    e^{-i\bm\Delta_\perp\cdot\bm x_\perp}
    F(-\bm\Delta_\perp^2).
\end{equation}
Here $x_\perp=|\bm x_\perp|$ after angular integration.

\begin{figure}[!htbp]
    \centering
    \includegraphics[width=0.48\linewidth]{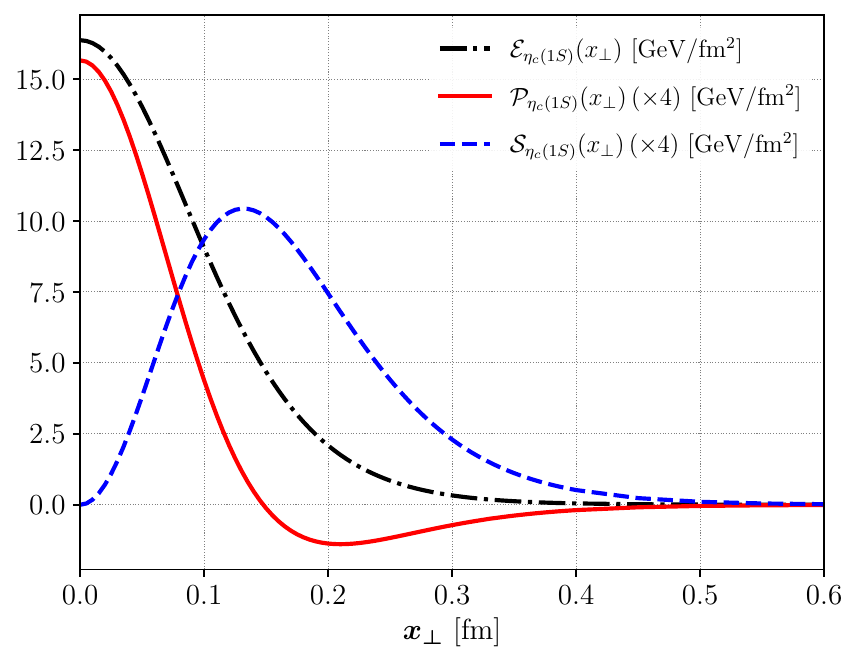}
    \includegraphics[width=0.48\linewidth]{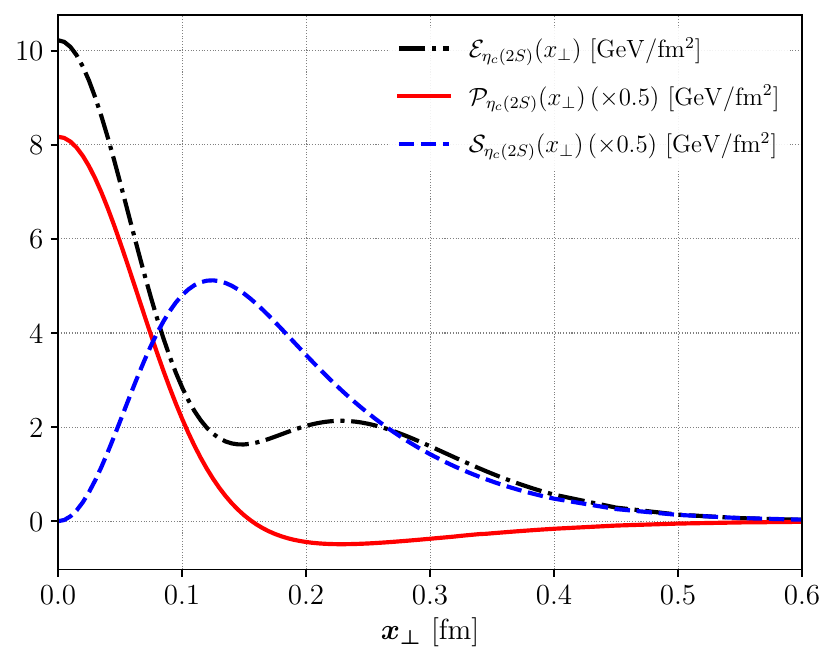}
    \includegraphics[width=0.48\linewidth]{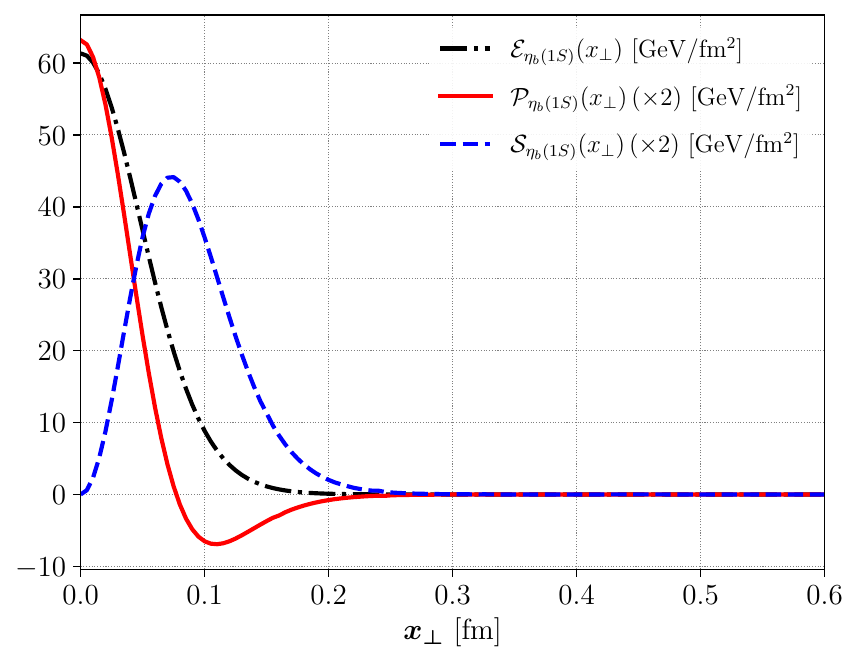}
    \includegraphics[width=0.48\linewidth]{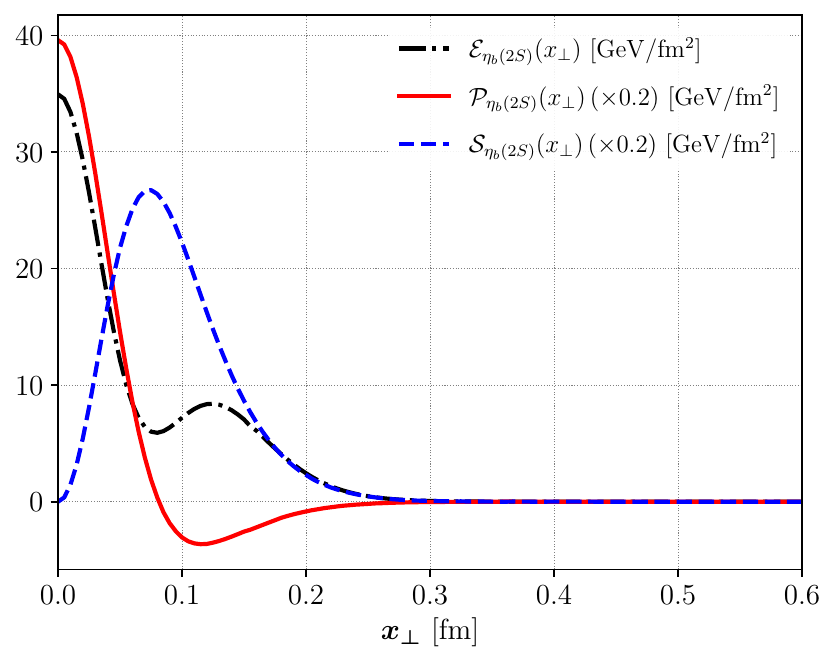}
    \caption{Transverse density profiles evaluated at
    $P^+=1$~GeV for charmonium (upper row) and bottomonium
    (lower row), with the physical $1S$ states on the left
    and $2S$ states on the right.
    Black dot-dashed, red solid, and blue dashed curves show
    $\mathcal E$, $\mathcal P$, and $\mathcal S$, respectively,
    in GeV/fm$^2$.
    For display, both the pressure and shear curves are
    multiplied by $4$, $0.5$, $2$, and $0.2$ for
    $\eta_c(1S)$, $\eta_c(2S)$, $\eta_b(1S)$, and
    $\eta_b(2S)$, respectively.}
    \label{fig:densities}
\end{figure}

{All density profiles in Fig.~\ref{fig:densities} are evaluated at the common value $P^+=1$~GeV.
Changing $P^+$ rescales $\mathcal E$ proportionally to $P^+$,
and $\mathcal P$ and $\mathcal S$ inversely with $P^+$,
without changing their transverse shapes or characteristic
length scales.}
The normalized $A$ profile is
$\mathcal E/P^+=\widetilde A$, whose integral equals
$A(0)=1$, whereas the integral of $\mathcal E$ equals $P^+$.
Its transverse mean-square moment is
$\langle x_\perp^2\rangle_A=4A'(0)
=(2/3)\langle r_A^2\rangle$, rather than the
three-dimensional mean-square slope radius itself.
The corresponding root-mean-square transverse sizes,
calculated from the mass radii in
Table~\ref{table:gff_radii}, are approximately
$0.156$ and $0.279$~fm for $\eta_c(1S,2S)$, and
$0.075$ and $0.139$~fm for $\eta_b(1S,2S)$.
These moments characterize the transverse spatial distribution
of plus momentum and should not be interpreted as moments
of a constituent position probability in a three-dimensional
rest-frame wave function.

Figure~\ref{fig:densities} shows a centrally concentrated,
monotonically decreasing $A$-density profile for each $1S$
state.
For $2S$, an additional shoulder or secondary maximum appears
at finite radius, in the region $x_\perp\simeq0.2$--$0.3$~fm
for charmonium and $0.1$--$0.15$~fm for bottomonium.
This is the coordinate-space manifestation of the radial
structure that also modifies the $2S$ mass form factor.
The redistribution toward larger $x_\perp$ contributes more
strongly to the second moment than to the normalization,
which accounts for the enhanced mass radius without changing
$A(0)$.
The more compact bottomonium profiles agree with the width
and radius comparisons above.

The stress profiles respond differently to radial excitation.
The pressure is positive in the central region and negative
outside a node near $x_0\simeq0.15$~fm for charmonium and
$0.08$~fm for bottomonium in both radial states.
The shear profiles vanish at the origin, develop a positive
maximum at finite radius, and decrease toward zero outside
the central region.
The relatively small change in the pressure-node position
from $1S$ to $2S$ is consistent with the modest change in
the mechanical radius, despite the much larger change
in $D(0)$.
Since all four states use the same $P^+$, their pressure
and shear magnitudes can be compared after removing the
panel-dependent display multipliers.
The displayed peak heights should therefore not be compared
directly across panels.

For regular transforms that decrease sufficiently rapidly,
integration by parts {in the pressure definition} gives
\begin{equation}\label{eq:2D-pressure-moments}
  \int d^2\bm x_\perp\,\mathcal P(x_\perp)=0,
  \qquad
  D(0)
  =
  2P^+\int d^2\bm x_\perp\,
  x_\perp^2\mathcal P(x_\perp).
\end{equation}
The first identity is the transverse von~Laue condition.
The second shows that the outer region of the pressure enters $D(0)$
with the additional weight $x_\perp^2$.
Note that a larger magnitude of $D(0)$ need not imply
a larger pressure-node position $x_0$ or D-term slope radius.
The positive central and negative outer regions in
Fig.~\ref{fig:densities} are consistent with this balance,
although a sign change alone is not a numerical verification
of the integrated condition or of local normal-force positivity.
Finally, the central density and the mechanical radius depend
on the full spacelike form factors.
Agreement of NR approximations only in the near-forward
interval does not by itself determine these quantities.

\section{Summary}\label{Sec:VI}

We have studied the GFFs of $\eta_c(1S,2S)$ and
$\eta_b(1S,2S)$ in the fBT-LFQM using spectroscopy-constrained
orthogonal mixtures of the two lowest HO basis states.
The fBT prescription applies the replacement of physical
meson masses by free-constituent invariant masses consistently
to both the constituent matrix elements and the Lorentz
prefactors inside the overlap integral.
{The three determinations of each form factor agree
in the full relativistic calculation.
The transverse-trace agreement follows after convolution with
the radial wave functions and requires the complete spin trace.
The independent extraction of $\bar c_q(t)$ gives
$\bar c_q(t)=0$ within the same fBT prescription, as shown in
Appendix~\ref{Sec:cbar-verification}.}
The equal quark and antiquark contributions give the total
valence form factors $A(t)=2A_q(t)$ and $D(t)=2D_q(t)$,
with $A(0)=1$ for every physical state.

We have derived complete analytic NR expressions through
$\mathcal O(v^2)$ for the diagonal HO contributions and
their interference, using $v=\beta/m_Q$ at fixed
$\rho=-t/\beta^2$.
The expansion retains the full kernel--overlap products,
including the Gaussian, Jacobian, and node-factor corrections.
In particular, the odd $v^{-1}$ term in
$\mathcal K_D^{(+-)}$ requires the overlap through
$\mathcal O(v^3)$ to obtain the complete $\mathcal O(v^2)$
form factor.
Current-component independence is preserved through this order.

The analytic results identify the radial origin of the leading
$D$-term within the present prescription.
The $\mathcal O(v^0)$ contribution cancels after integration
for the pure Gaussian $1S$ state but survives for the $2S$
basis state.
The $D$-term interference first contributes at $\mathcal O(v^2)$.
Consequently, a nonzero $2S$ admixture held fixed as $v\to0$
leaves an $\mathcal O(v^0)$ $D$-term in the physical ground state.
This Gaussian cancellation is not a general consequence of
a nodeless momentum-space wave function.
Mixing also affects the forward slopes.
In the adopted convention, interference lowers the ground-state
$A'(0)$ and raises the excited-state slope relative to the
diagonal contributions alone.
For $D(t)$, interference contributes to $D(0)$ but has a
vanishing forward slope through $\mathcal O(v^2)$.
The forward normalization and slope therefore distinguish
different contributions to the form factors.

For the physical mixtures with $\theta=12^\circ$, the full
relativistic calculation gives the total valence values
$D_{\eta_c(1S)}(0)=-0.370$,
$D_{\eta_c(2S)}(0)=-2.088$,
$D_{\eta_b(1S)}(0)=-0.201$, and
$D_{\eta_b(2S)}(0)=-1.649$.
Radial excitation substantially increases $|D(0)|$,
while bottomonium has smaller forward magnitudes than
charmonium for the corresponding radial states.
The charmonium mass radii are numerically close to the
BLFQ results despite substantially smaller
$|D(0)|$~\cite{Xu:2024hfx,Hu:2024edc}.
Our $\eta_c(1S)$ forward $D$-term is numerically closer
to the contact-interaction result obtained from coupled
Dyson--Schwinger and Bethe--Salpeter equations,
although its magnitude remains smaller~\cite{Sultan:2024hep}.

Comparisons with the full calculation use the same
wave-function parameters and mixing conventions at each
order of the NR approximation.
The analytic NLO expressions reproduce $A'(0)$ at the
few-percent level for charmonium, with closer agreement
for bottomonium.
The $D$-term provides a more sensitive test.
At NLO, $|D(0)|$ is overestimated while $|D'(0)|$ is
underestimated, so the normalized slope $D'(0)/D(0)$
is smaller than in the full calculation.
The suppressed leading ground-state $D$-term makes its
relative correction particularly large.
The numerically evaluated $\mathcal O(v^4)$ corrections
bring the approximations closer to the full results in
the near-forward region and around the physical values
of $v$ examined here.
The fixed-$\rho$ comparisons further show the stronger
higher-order effects near the charmonium value of $v$.
Thus, NR accuracy must be assessed separately for
$A'(0)$, $D(0)$, and $D'(0)$ rather than inferred from
the expansion parameter alone.

The radii and transverse densities reveal complementary
effects of heavy-quark flavor and radial excitation.
Bottomonium is more compact than charmonium for both
radial states.
Radial excitation increases the mass radius substantially
more than the $D$-term slope radius or the mechanical radius.
For both flavors, the mass radius is the smallest of the
three gravitational radii for $1S$ and the largest for $2S$,
while the mechanical radius remains smaller than the
$D$-term slope radius.
The transverse $A$ profiles develop an additional shoulder
or secondary maximum for $2S$, whereas the pressure-node
positions change much less under radial excitation.
The pressure profiles have positive central and negative
outer regions.
These results distinguish changes in the overall $D$-term
magnitude from changes in the spatial extent of the stress
distributions.
They also emphasize that central densities and mechanical
radii require the full spacelike form factors, not only
their near-forward NR behavior.

The present results provide a common description of the
relativistic form factors and their NR limits within the
valence constituent-current framework.
Extensions to unequal-mass systems such as $B_c$ require
an expansion about $x=m_1/(m_1+m_2)$ and a corresponding
treatment of asymmetric kinematics and wave-function overlaps.
Other directions include orbital and spin channels such as
$\chi_{c0}$ and $J/\psi$. { The investigation in the $\Delta^+ \neq 0$ frame would also deserve further analysis in particular for the computation in the timelike ($t=\Delta^2 >0$) region.}
%%%%%%%%%%%%%%%%%%%%%%%%%%%%%%%%%%%%%%%%%%%%%%%%%%%

%%%%%%%%%%%%%%%%%%%%%%%%%%%%%%%%%%%%%%%%%%%%%%%%%%% 
\section*{Acknowledgements}
The present work was supported by the National Research Foundation of Korea (NRF) grant funded by the Korea government (MSIT) under Grant Nos. RS-2026-25475296 (H.-M. Choi) and RS-2023-00210298 (H.-D. Son). C.-R. Ji was supported in part by the U.S. Department of Energy (Grant No. DE-FG02-03ER41260). The National Energy Research Scientific Computing Center (NERSC), supported by the Office of Science of the U.S. Department of Energy under Contract No. DE-AC02-05CH11231, is also acknowledged.
%%%%%%%%%%%%%%%%%%%%%%%%%%%%%%%%%%%%%%%%%%%%%%%
\appendix
\providecommand{\revsixteen}[1]{{#1}}
\counterwithin*{equation}{section}
\renewcommand{\theequation}{\Alph{section}\arabic{equation}}
\setcounter{equation}{0}

\Needspace{10\baselineskip}
\section{Verification of $\bar c_q(t)=0$ within the fBT prescription}
\label{Sec:cbar-verification}

{
With the metric term retained in Eq.~\eqref{eq:GFFdecomp},
its $(+-)$ projection gives
\begin{equation}\label{eq:cbar-physical-extraction}
  \bar c_q(t)
  =\frac{\langle T_q^{+-}\rangle
  -\Gamma_A^{+-}A_q^{(++)}(t)
 -\Gamma_D^{+-}D_q^{(-i)}(t)}
 {4M^2}.
\end{equation}
Here $a^\pm=a^0\pm a^3$ gives $g^{+-}=2$.
Since $g^{++}=g^{-i}=0$ and $\Gamma_D^{++}=0$, the $(++)$
component determines $A_q$, and the contracted $(-i)$ component
determines $D_q$ after the $A_q$ subtraction, without a
metric-term contribution.
The fBT prescription of Eq.~\eqref{eq:self-consistent} applies
this linear extraction inside the overlap using the unreduced
Lorentz structures of Eq.~\eqref{eq:LorentzStruct}, including
the mass replacement in the metric prefactor,
\begin{equation}\label{eq:cbar-prefactor}
  \Gamma_{\bar c}^{\mu\nu}
  \equiv[2M^2g^{\mu\nu}]_{\rm BT}
  =2M_0^2g^{\mu\nu},
  \qquad
  \Gamma_{\bar c}^{+-}=4M_0^2.
\end{equation}
The fBT extraction specified by
Eqs.~\eqref{eq:cbar-physical-extraction} and~\eqref{eq:cbar-prefactor}
gives
\begin{equation}\label{eq:cbar-convolution}
  \bar c_q(t)
  =\int_0^1 dx\int\frac{d^2\bm k_\perp}{2(2\pi)^3}\,
    \Phi'^{*}(x,\bm k'_\perp)\,
    \Phi(x,\bm k_\perp)\,
    \mathcal K_{\bar c}^{(+-)}
    (x,\bm k_\perp,\bm\Delta_\perp),
\end{equation}
where
\begin{equation}\label{eq:cbar-kernel}
  \mathcal K_{\bar c}^{(+-)}
  \equiv\frac{P^+S_q^{+-}-\Gamma_A^{+-}\mathcal K_A
  -\Gamma_D^{+-}\mathcal K_D^{(-i)}}{4M_0^2}
  =\frac{\bm\Delta_\perp^2}{4M_0^2}
    \left[\mathcal K_D^{(+-)}-\mathcal K_D^{(-i)}\right].
\end{equation}
Here $\mathcal K_A=\mathcal K_A^{(++)}$ and
$\mathcal K_D^{(-i)}$ are the corresponding extraction kernels
given in Eqs.~\eqref{eq:Akernels} and~\eqref{eq:Dkernels},
respectively.
The factor $P^+$ originates from the measure in
Eq.~\eqref{eq:convolution}.
The second equality uses Eq.~\eqref{eq:Dkernels} with
$\Gamma_D^{+-}=\bm\Delta_\perp^2=-t>0$.

The cancellation can be shown directly from
Table~\ref{tab:trace}.
The spectator relation gives
\begin{equation}\label{eq:cbar-spectator}
  \bm p_{s\perp}\cdot\bm\Delta_\perp
  =2\bm k_\perp\cdot\bm\Delta_\perp+\bm\Delta_\perp^2
  =x(M_-^2+\bm\Delta_\perp^2).
\end{equation}
Together with $p_1^+=xP^+$ and $\Delta_i e^i=0$,
the $(+-)$ and $(-i)$ rows of Table~\ref{tab:trace} give
\begin{align}\label{eq:cbar-proportionality}
  S_q^{-i}\Delta_i&=\omega S_q^{+-},
  \nonumber\\
  \Gamma_A^{-i}\Delta_i&=\omega\Gamma_A^{+-},
  \nonumber\\
  \Gamma_D^{-i}\Delta_i&=\omega\Gamma_D^{+-},
\end{align}
where
\begin{equation}\label{eq:cbar-omega}
  \omega\equiv\frac{p_s^i\Delta_i}{2p_1^+}
  =\frac{M_-^2+\bm\Delta_\perp^2}{2P^+}.
\end{equation}
Substitution into the $(-i)$ extraction of
Eq.~\eqref{eq:Dkernels} yields
\begin{equation}\label{eq:cbar-D-identity}
  \mathcal K_D^{(-i)}
  =\frac{P^+S_q^{+-}-\Gamma_A^{+-}\mathcal K_A}
 {\Gamma_D^{+-}}
  =\mathcal K_D^{(+-)}.
\end{equation}
This is a pointwise identity between the reduced kernels,
with the $(-i)$ kernel defined by continuity at $\omega=0$.
Therefore, $\mathcal K_{\bar c}^{(+-)}=0$, and
Eq.~\eqref{eq:cbar-convolution} gives $\bar c_q(t)=0$.
This result follows from the independently evaluated $(+-)$
and contracted $(-i)$ spin traces for any convergent radial
overlap, without an NR expansion.

The transverse trace, with
$\Gamma_D^{ii}=-\bm\Delta_\perp^2/2$ and
$\Gamma_{\bar c}^{ii}=-4M_0^2$, gives
\begin{equation}\label{eq:cbar-trace-kernel}
  \mathcal K_{\bar c}^{(ii)}
  \equiv\frac{P^+S_q^{ii}-\Gamma_A^{ii}\mathcal K_A
   -\Gamma_D^{ii}\mathcal K_D^{(-i)}}{-4M_0^2}
  =\frac{\bm\Delta_\perp^2}{8M_0^2}
    \left[\mathcal K_D^{(ii)}-\mathcal K_D^{(-i)}\right].
\end{equation}
Numerical evaluation of Eq.~\eqref{eq:cbar-convolution}
with $\mathcal K_{\bar c}^{(ii)}$ also gives
$\bar c_q^{(ii)}(t)=0$ within numerical accuracy
for the radial $S$-wave overlaps considered here.
This result is obtained with the fBT prescription applied
at the integrand level.
}
{ It confirms the consistency of the fBT prescribed formulation
regarding the independence of tensor components for the extraction of
GFFs.}

\section{Heavy quarkonium in the nonrelativistic limit: explicit derivation}
\label{Sec:NRQCD}
%%%%%%%%%%%%%%%%%%%%%%%%%%%%%%%%%%%%%%%%%%%%%%%%%%%%%%%%%%%%%%%%%%%%%%

This appendix gives the detailed derivation of the analytic results in
Sec.~\ref{Sec:NRQCD-summary}, using the extraction kernels defined in
Eqs.~\eqref{eq:Akernels} and~\eqref{eq:Dkernels}.
Section~\ref{Sec:NRQCD-rad} defines the radial scales and overlap moments.
Section~\ref{Sec:NRQCD-kin} gives the peak-evaluated kinematic entries of
Table~\ref{tab:NR-2}. Section~\ref{Sec:NRQCD-W} expands the radial overlap
common to both extractions, and Secs.~\ref{Sec:NRQCD-A}
and~\ref{Sec:NRQCD-D} combine it with the respective kernels.

\subsection{Internal scales and the radial wave function}
\label{Sec:NRQCD-rad}

Using the equal-mass kinematics of Sec.~\ref{Sec:II} and the power
counting of Sec.~\ref{Sec:NRQCD-summary}, we expand the constituent
energy $E=E_1=E_2$ as
$E=m_Q+\vec{k}^{\,2}/(2m_Q)+\mathcal O(m_Qv^4)$ to obtain
\begin{equation}\label{eq:NR-LO}
  x=\frac12+\frac{k_z}{2m_Q}+\mathcal{O}(v^3),
  \qquad
  M_0=2m_Q+\frac{\vec{k}^{\,2}}{m_Q}
       +\mathcal{O}(m_Qv^4).
\end{equation}
For characteristic internal momenta $|\vec{k}|\sim\beta$, the first
relation implies $|x-\tfrac12|\sim\beta/(2m_Q)=\mathcal O(v)$.

The term $\vec k^{\,2}/m_Q=\vec k^{\,2}/(2\mu)$, with
$\mu=m_Q/2$, is the leading internal kinetic energy. It must be
retained in the free-constituent invariant masses. The exact relation
$M_-^2=4(\vec k'^{\,2}-\vec k^{\,2})$ and its recoil scaling are used
in Sec.~\ref{Sec:NRQCD-kin}.

\emph{Longitudinal variable and measure.}
For the exact change of variables in Eq.~\eqref{eq:NR-KW-convolution},
$x$ is given by Eq.~\eqref{eq:kz}, with $k_z$ and $\bm k_\perp$
independent. The initial- and final-state radial functions contain the
Jacobians $M_0/[4x(1-x)]$ and $M'_0/[4x(1-x)]$, respectively.
The latter is evaluated at fixed $\bm k'_\perp$, not as a total
derivative along the $x$-dependent recoil shift.
Factoring out the square roots of these Jacobians gives
\begin{equation}\label{eq:NR-measure}
  \phi^{*}(x,\bm k'_\perp)\,\phi(x,\bm k_\perp)\,dx
  =\sqrt{\frac{M'_0}{M_0}}\,
   \widetilde\phi^{*}(\vec k\,')\,
   \widetilde\phi(\vec k)\,dk_z.
\end{equation}
This yields the overlap factor in Eq.~\eqref{eq:NR-KW-overlap}
and the convolution in Eq.~\eqref{eq:NR-KW-convolution}.
The exact relation $k_z=(x-\tfrac12)M_0$ maps $0<x<1$ onto the
entire real line, so no endpoint approximation enters.
The HO radial functions are real, and complex-conjugation symbols
are therefore omitted below.

\emph{$\phi_{1S}$ state.}
The normalized Gaussian $\widetilde\phi_{1S}$ obtained from
Eq.~\eqref{eq:HOWF2} gives $\langle\vec{k}^{\,2}\rangle_{t=0}=\tfrac32\beta^2$ and, by
isotropy, $\langle k_z^2\rangle_{t=0}=\tfrac12\beta^2$.
At leading order, the recoil relations reduce to
$\bm k'_\perp=\bm k_\perp+\bm\Delta_\perp/2$,
$k'_z=k_z$, and $\sqrt{M'_0/M_0}=1$, while $k_z$ remains an independent
integration variable. Using
$\bm k_\perp^2+\bm k_\perp'^2
=2(\bm k_\perp+\bm\Delta_\perp/4)^2+\bm\Delta_\perp^2/8$,
the leading overlap in Eq.~\eqref{eq:NR-KW-overlap} becomes
\begin{equation}\label{eq:NR-WFproduct-Gaussian}
  \mathcal W_{G,0}\equiv
  \Bigl[\sqrt{M'_0/M_0}\;\widetilde\phi\,'_{1S}\widetilde\phi_{1S}\Bigr]_{\rm LO}
  =\frac{16\pi^{3/2}}{\beta^3}\,G(\rho)\,
  \exp\!\left[
    -\frac{(\bm{k}_\perp+\tfrac14\bm{\Delta}_\perp)^2+k_z^2}{\beta^2}
  \right],
\end{equation}
with $G(\rho)$ as introduced after Eq.~\eqref{eq:NR-Aq-summary}.
Corrections from the exact transverse shift, longitudinal recoil,
and Jacobian are evaluated in Sec.~\ref{Sec:NRQCD-W}.
We define the normalized average with this leading weight,
\begin{equation}\label{eq:NR-avg-def}
  \langle f\rangle_{\phi_{1S}}\equiv
  \frac{\displaystyle\int\frac{dk_z\,d^2\bm{k}_\perp}{2(2\pi)^3}\,
    f\,\mathcal W_{G,0}}
  {\displaystyle\int\frac{dk_z\,d^2\bm{k}_\perp}{2(2\pi)^3}\,\mathcal W_{G,0}}
  =\frac1{\pi^{3/2}}\int d^3\vec u\,e^{-\vec u^{\,2}}f,
\end{equation}
with $\vec u=(\bm k_\perp+\bm\Delta_\perp/4,k_z)/\beta$. In the last
integral, $f$ is evaluated at $k_z=\beta u_z$ and
$\bm k_\perp=\beta\bm u_\perp-\bm\Delta_\perp/4$.
The denominator equals $G(\rho)$, so a convolution with
$\mathcal W_{G,0}$ equals $G(\rho)\langle f\rangle_{\phi_{1S}}$.
The brackets $\langle\cdots\rangle_{\phi_{1S}}$, and the
$\phi_{2S}$ brackets defined below, refer to the leading overlap
weights, not to the full finite-$v$ overlaps. Higher-order overlap
corrections are included explicitly in the coefficient products.
%\endgroup

Introducing $\bm{l}_\perp=\bm{k}_\perp+\tfrac14\bm{\Delta}_\perp$,
Gaussian integration gives
\begin{equation}\label{eq:NR-moments}
  \begin{aligned}
    \langle\bm{k}_\perp\rangle_{\phi_{1S}}
      &=-\tfrac14\bm{\Delta}_\perp,
    &
    \langle\bm{k}_\perp\!\cdot\!\bm{\Delta}_\perp\rangle_{\phi_{1S}}
      &=-\tfrac14\bm{\Delta}_\perp^2,
    \\
    \langle\bm{k}_\perp^2\rangle_{\phi_{1S}}
      &=\beta^2+\tfrac1{16}\bm{\Delta}_\perp^2,
    &
    \langle k_z^2\rangle_{\phi_{1S}}
      &=\tfrac12\beta^2,
  \end{aligned}
\end{equation}
together with $\langle x-\tfrac12\rangle_{\phi_{1S}}=0$ and
$\langle(x-\tfrac12)^2\rangle_{\phi_{1S}}=\beta^2/(8m_Q^2)+\mathcal{O}(v^4)$,
from $x-\tfrac12=k_z/M_0$.

\emph{$\phi_{2S}$ state.}
The three-momentum radial function $\widetilde\phi_{2S}$ obtained
from Eq.~\eqref{eq:HOWF2} is rotationally symmetric in $\vec{k}$,
with a radial node at
$|\vec{k}|=\sqrt{3/2}\,\beta$ and the forward moment
$\langle\vec{k}^{\,2}\rangle_{t=0}=\tfrac72\beta^2$.

For the leading overlap, write $\vec{l}=(\bm{l}_\perp,k_z)$ using the
shifted transverse momentum defined above. Since $\widetilde{\phi}_{2S}$ and
$\widetilde{\phi}_{1S}$ differ only by the factor
$(2\vec{k}^{\,2}-3\beta^2)/(\sqrt6\,\beta^2)$, the $\phi_{2S}$ overlap
factorizes as
\begin{equation}\label{eq:NR-nodeproduct}
  \Bigl[\sqrt{M'_0/M_0}\;\widetilde\phi\,'_{2S}\widetilde\phi_{2S}\Bigr]_{\rm LO}
  =\Pi_2\,\mathcal W_{G,0},
\end{equation}
where
\begin{equation}\label{eq:NR-nodeproduct2}
  \Pi_2(\vec l,\bm\Delta_\perp)
  \equiv
  \left[
    \frac{(2\vec k'^{\,2}-3\beta^2)
          (2\vec k^{\,2}-3\beta^2)}{6\beta^4}
  \right]_{\rm LO}
  =
  \frac{
    (2\vec l^{\,2}+\bm\Delta_\perp^2/8-3\beta^2)^2
    -(\bm l_\perp\!\cdot\!\bm\Delta_\perp)^2
  }{6\beta^4}.
\end{equation}
The second equality in Eq.~\eqref{eq:NR-nodeproduct2} uses
$\bm{k}_\perp=\bm{l}_\perp-\tfrac14\bm{\Delta}_\perp$,
$\bm{k}'_\perp=\bm{l}_\perp+\tfrac14\bm{\Delta}_\perp$,
and $k'_z=k_z$ at this order.
The Gaussian average $\langle\Pi_2\rangle_{\phi_{1S}}=P_2(\rho)$
reproduces the recoil polynomial introduced after
Eq.~\eqref{eq:NR-Aq-summary}.
The normalized leading $\phi_{2S}$ overlap averages can therefore
be expressed through the $\phi_{1S}$ average as
\begin{equation}\label{eq:NR-avg-2S}
  \langle f\rangle_{\phi_{2S}}
  =\frac{\langle f\,\Pi_2\rangle_{\phi_{1S}}}{P_2(\rho)}.
\end{equation}
Correspondingly, the unnormalized leading $\phi_{2S}$ overlap is
\begin{equation}\label{eq:NR-P2def}
  \int\frac{dk_z\,d^2\bm{k}_\perp}{2(2\pi)^3}\,\Pi_2\,\mathcal W_{G,0}
  =G(\rho)\,P_2(\rho).
\end{equation}
A convolution with this leading weight is therefore
$G(\rho)P_2(\rho)\langle f\rangle_{\phi_{2S}}$.

Gaussian integration gives
\begin{equation}\label{eq:NR-moments-2S}
  \begin{aligned}
    \langle\bm{k}_\perp\rangle_{\phi_{2S}}
      &=-\tfrac14\bm{\Delta}_\perp,
    &
    \langle\bm{k}_\perp\!\cdot\!\bm{\Delta}_\perp\rangle_{\phi_{2S}}
      &=-\tfrac14\bm{\Delta}_\perp^2,
    \\
    \langle\bm{k}_\perp^2\rangle_{\phi_{2S}}
      &=\frac{\bm{\Delta}_\perp^2}{16}+\frac{\beta^2}{P_2(\rho)}
      \left[\frac73-\frac{\rho}{12}+\frac{\rho^2}{384}\right],
    &
    \langle k_z^2\rangle_{\phi_{2S}}
      &=\frac{\beta^2}{P_2(\rho)}
      \left[\frac76+\frac{\rho^2}{768}\right],
  \end{aligned}
\end{equation}
together with
$\langle(x-\tfrac12)^2\rangle_{\phi_{2S}}
=\langle k_z^2\rangle_{\phi_{2S}}/(4m_Q^2) +\mathcal{O}(v^4)$.

Two properties of this overlap must be distinguished.
At fixed $\bm{\Delta}_\perp$ and $k_z$, the product
$\Pi_2\,\mathcal W_{G,0}$
is invariant under $\bm{l}_\perp\to-\bm{l}_\perp$, so odd moments of
$\bm{l}_\perp$ vanish as for $\phi_{1S}$, giving the first two moments above.
Reflection symmetry does not imply isotropy at finite recoil. The term
$-(\bm{l}_\perp\!\cdot\!\bm{\Delta}_\perp)^2$
in Eq.~\eqref{eq:NR-nodeproduct2} distinguishes the direction of
$\bm{\Delta}_\perp$ from the orthogonal transverse direction,
so that $\langle\bm{k}_\perp^2\rangle_{\phi_{2S}}$ and
$\langle k_z^2\rangle_{\phi_{2S}}$ acquire the $\rho$ dependence shown above
and reduce to $\tfrac73\beta^2$ and $\tfrac76\beta^2$ only in the forward
limit.

\subsection{Reduction of the spin traces and Lorentz prefactors}
\label{Sec:NRQCD-kin}

We evaluate the entries of Table~\ref{tab:trace} at the reference
point $x=1/2$, where
$p_1^+=P^+/2$,
$\bm k'_\perp=\bm k_\perp+\bm\Delta_\perp/2$, and
$k_z=k'_z=0$.
The resulting expressions, collected in Table~\ref{tab:NR-2},
illustrate the kinematic reduction.
They do not replace the complete NLO integrands, which retain
$x=x(k_z,\bm k_\perp)$ before expansion.

For general $x$, the spectator shift gives
\begin{equation}\label{eq:NR-Mratio}
  \frac{M_-^2}{\bm\Delta_\perp^2}
  =
  \frac{2\bm k_\perp\cdot\bm\Delta_\perp
        +(1-x)\bm\Delta_\perp^2}
       {x\,\bm\Delta_\perp^2}
  =
  \mathcal O(1),
\end{equation}
where the counting follows from
$|\bm\Delta_\perp|\sim|\bm k_\perp|\sim m_Qv$ and
$x=\tfrac12+\mathcal O(v)$.
Although $M_-^2=\mathcal O(m_Q^2v^2)$, its ratio to
$\bm\Delta_\perp^2$ is therefore not suppressed.
Likewise, the term
$(M_-^2)^2/\bm\Delta_\perp^2=\mathcal O(m_Q^2v^2)$
in $\Gamma_A^{ij}$ is of the same order as
$\Delta^i\Delta^j$ and must be retained.

In the full convolution, the same counting gives
$M'_0/M_0-1=\mathcal O(v^2)$.
Together with $k'_z=k_zM'_0/M_0$ and $k_z\sim m_Qv$,
this implies
$k'_z-k_z=\mathcal O(m_Qv^3)$ and
$(k_z'^2-k_z^2)/\beta^2=\mathcal O(v^2)$.
The longitudinal recoil therefore contributes to the NLO
overlap expansion.

At $x=1/2$, the common inverse normalization factor and the
constituent minus-momentum sum are
\begin{equation}\label{eq:NR-spinor-norm}
  \begin{aligned}
    \left.
    \frac{1}
         {\sqrt{m_Q^2+\bm k_\perp^2}
          \sqrt{m_Q^2+\bm k_\perp^{\prime 2}}}
    \right|_{x=1/2}
    &=
    \frac{1}{m_Q^2}
    \left[
      1-\frac{\bm k_\perp^2+\bm k_\perp^{\prime 2}}
              {2m_Q^2}
      +\mathcal O(v^4)
    \right],
    \\[1ex]
    \left.(p_1+p'_1)^-\right|_{x=1/2}
    &=
    \frac{
      4m_Q^2
      +2\left[
        \bm k_\perp^2
        +(\bm k_\perp+\bm\Delta_\perp)^2
      \right]}
      {P^+}.
  \end{aligned}
\end{equation}
The second line follows exactly from the on-shell condition.
Multiplying the inverse normalization by the spin-overlap
numerator gives
\be
  \frac{m_Q^2+\bm k_\perp\cdot\bm k'_\perp}
       {\sqrt{m_Q^2+\bm k_\perp^2}
        \sqrt{m_Q^2+\bm k_\perp^{\prime 2}}}
  =
  1-\frac{(\bm k'_\perp-\bm k_\perp)^2}{2m_Q^2}
  +\mathcal O(v^4).
\ee

The auxiliary quantity $S_c$ of Table~\ref{tab:trace} has the
exact fixed-$x$ form
\begin{equation}\label{eq:NR-Sc}
  \begin{aligned}
    \left.S_c\right|_{x=1/2}
    &=
    m_Q^4
    +m_Q^2\left(
      \bm k_\perp^2
      +\bm k_\perp\cdot\bm k'_\perp
      +\bm k_\perp\cdot\bm\Delta_\perp
      +\frac12\bm\Delta_\perp^2
    \right)
    \\
    &\quad
    +(\bm k_\perp\cdot\bm k'_\perp)
     (\bm k_\perp^2+\bm k_\perp\cdot\bm\Delta_\perp)
    +\frac12\chi^2.
  \end{aligned}
\end{equation}
The last two terms are of $\mathcal O(m_Q^4v^4)$.
After multiplication by the inverse normalization and division
by $\Gamma_D^{+-}=\bm\Delta_\perp^2$, terms of this order can
contribute to $\mathcal K_D$ at $\mathcal O(v^2)$.
They are therefore retained in the complete kernel expansion
of Sec.~\ref{Sec:NRQCD-D}.

The invariant-mass combinations at the same reference point are
\begin{align}
  M_+^{2,\rm NR}
  &\equiv
  \left.M_+^2\right|_{x=1/2}
  =
  8m_Q^2+4(\bm k_\perp^2+\bm k_\perp^{\prime 2}),
  \label{eq:NR-Mplus}\\
  M_-^{2,\rm NR}
  &\equiv
  \left.M_-^2\right|_{x=1/2}
  =
  4\bm k_\perp\cdot\bm\Delta_\perp+\bm\Delta_\perp^2.
  \label{eq:NR-Mminus-equal}
\end{align}
These relations are exact at $x=1/2$.

Keeping relative $\mathcal O(v^2)$ corrections in the spin
traces gives the entries of Table~\ref{tab:NR-2}, written in
terms of
\begin{equation}\label{eq:NR-Spm-bracket}
  \begin{aligned}
    S_{\rm NR}^{++}
    &\equiv
    1-\frac{\bm\Delta_\perp^2}{8m_Q^2},
    \\
    S_{\rm NR}^{+-}
    &\equiv
    4m_Q^2
    +4\bm k_\perp\cdot\bm k'_\perp
    +2\bm k_\perp\cdot\bm\Delta_\perp
    +\frac32\bm\Delta_\perp^2.
  \end{aligned}
\end{equation}
The Lorentz prefactors follow by inserting
Eqs.~\eqref{eq:NR-Mplus} and~\eqref{eq:NR-Mminus-equal}
into Table~\ref{tab:trace}.
In particular, $\Gamma_A^{++}=2(P^+)^2$ and
$\Gamma_D^{++}=0$ remain exact.

The terms proportional to $e^i$ contribute at relative
$\mathcal O(v^2)$ in the $(+i)$, $(-i)$, and $(ij)$ spin traces.
Their momentum-transfer contractions vanish, while
$p_s^ie^i=2\chi$ gives
\be
  \left.S_q^{ii}\right|_{e,\,x=1/2}
  =
  \frac{\chi^2}{P^+m_Q^2}
  \left[1+\mathcal O(v^2)\right].
\ee
All uncontracted entries in Table~\ref{tab:NR-2} retain these
terms.

\begin{table*}[!htbp]
\caption{\label{tab:NR-2}%
Reduction of Table~\ref{tab:trace} at $x=1/2$, retaining relative
$\mathcal O(v^2)$ corrections in the spin traces.
The quantities $M_\pm^{2,\rm NR}$ and $S_{\rm NR}^{++,+-}$
are defined in Eqs.~\eqref{eq:NR-Mplus},
\eqref{eq:NR-Mminus-equal}, and~\eqref{eq:NR-Spm-bracket}.
The definitions of $p_s^i$, $e^i$, and $\chi$ are given in
Table~\ref{tab:trace}.
All $e^i$ terms are retained in the uncontracted entries.
This fixed-$x$ reduction does not replace the complete NLO
kernel expansion.}
\small
\begin{tabular}{
  c@{\hskip 0.3cm}
  c@{\hskip 0.3cm}
  c@{\hskip 0.3cm}
  c}
\hline\hline\\[-1.5ex]
$(\mu\nu)$
& $S_q^{\mu\nu}$
& $\Gamma_A^{\mu\nu}$
& $\Gamma_D^{\mu\nu}$\\[1.5ex]
\hline\\[-1ex]

$(++)$
& $P^+S_{\rm NR}^{++}$
& $2(P^+)^2$
& $0$\\[3ex]

$(+-)$
& $\dfrac{S_{\rm NR}^{+-}}{P^+}$
& $M_+^{2,\rm NR}+M_-^{2,\rm NR}+\bm\Delta_\perp^2$
& $\bm\Delta_\perp^2$\\[3ex]

$(+i)$
& $p_s^iS_{\rm NR}^{++}
   +\dfrac{\chi e^i}{4m_Q^2}$
& $P^+\Delta^i
   \left(1+\dfrac{M_-^{2,\rm NR}}{\bm\Delta_\perp^2}\right)$
& $0$\\[3ex]

$(-i)$
& $\dfrac{p_s^iS_{\rm NR}^{+-}+\chi e^i}{(P^+)^2}$
& $\dfrac{
    \Delta^i(\bm\Delta_\perp^2+M_-^{2,\rm NR})
    [M_+^{2,\rm NR}+M_-^{2,\rm NR}+\bm\Delta_\perp^2]}
    {2\bm\Delta_\perp^2P^+}$
& $\dfrac{\Delta^i(\bm\Delta_\perp^2+M_-^{2,\rm NR})}
          {2P^+}$\\[3ex]

$(ij)$
& $\dfrac{p_s^ip_s^j}{P^+}S_{\rm NR}^{++}
   +\dfrac{\chi(p_s^ie^j+p_s^je^i)}{4P^+m_Q^2}$
& $\dfrac12
   \left[
     \Delta^i\Delta^j
     \left(1+\dfrac{2M_-^{2,\rm NR}}{\bm\Delta_\perp^2}\right)
     -g^{ij}\dfrac{(M_-^{2,\rm NR})^2}{\bm\Delta_\perp^2}
   \right]$
& $\dfrac12
   (\Delta^i\Delta^j+g^{ij}\bm\Delta_\perp^2)$\\[3ex]

\hline\hline
\end{tabular}
\end{table*}

\subsection{Expansion of the overlap}
\label{Sec:NRQCD-W}

For each radial contribution, the overlap $\mathcal W$ of
Eq.~\eqref{eq:NR-KW-overlap} is common to the $A_q$ and $D_q$
extractions.
We derive its coefficients here for use in
Secs.~\ref{Sec:NRQCD-A} and~\ref{Sec:NRQCD-D}.
The leading correction odd in $k_z$ follows from the exact
relation $x-\tfrac12=k_z/M_0$, with
$M_0=2\sqrt{m_Q^2+\vec k^{\,2}}$.
The transverse spectator shift therefore differs from its
leading value $\bm\Delta_\perp/2$ by
\begin{equation}
  \left(\frac12-x\right)\bm\Delta_\perp
  =
  -\frac{k_z}{M_0}\bm\Delta_\perp
  =
  -\frac{k_z}{2m_Q}\bm\Delta_\perp
  +\mathcal O(\beta v^3).
\end{equation}
Consequently,
\begin{equation}\label{eq:NR-kprime-exp}
  \bm k_\perp^{\prime 2}
  =
  \left(\bm k_\perp+\frac12\bm\Delta_\perp\right)^2
  -\frac{k_z}{m_Q}\,
   \bm\Delta_\perp\cdot
   \left(\bm k_\perp+\frac12\bm\Delta_\perp\right)
  +\mathcal O(\beta^2v^2).
\end{equation}
For the pure Gaussian $1S$ overlap, exponentiating this
recoil correction gives
\begin{equation}\label{eq:NR-overlap-exp}
  \mathcal W^{(1S)}
  =
  \mathcal W_{G,0}
  \left[
    1+
    \frac{k_z
      \bigl(M_-^{2,\rm NR}+\bm\Delta_\perp^2\bigr)}
         {8m_Q\beta^2}
    +\mathcal O(v^2)
  \right],
\end{equation}
where $\mathcal W_{G,0}$ is the leading Gaussian weight of
Eq.~\eqref{eq:NR-WFproduct-Gaussian}.
Corrections from the Jacobian factor $\sqrt{M'_0/M_0}$
and longitudinal recoil first contribute at $\mathcal O(v^2)$.

Using the centered variable $\vec u$ of
Eq.~\eqref{eq:NR-avg-def}, we choose $\bm\Delta_\perp$
along the positive $x$ axis and write
\begin{equation}\label{eq:NR-D-scaled-vars}
  \lambda=\frac{|\bm\Delta_\perp|}{\beta}=\sqrt\rho,
  \qquad
  \kappa^2=\frac{\vec k^{\,2}}{\beta^2}
  =
  \left(u_x-\frac{\lambda}{4}\right)^2+u_y^2+u_z^2.
\end{equation}
The expansion is taken at fixed $\vec u$ and $\lambda$.

Retaining the exact longitudinal relation
$k'_z=k_zM'_0/M_0$ together with the transverse spectator
shift gives
\begin{equation}\label{eq:NR-D-recoil-cubic}
  \begin{aligned}
    \kappa'^2
    \equiv\frac{\vec k'^{\,2}}{\beta^2}
    &=
    \kappa_0'^2+v\epsilon_1+v^2\epsilon_2
    +v^3\epsilon_3+\mathcal O(v^4),
    \\
    \kappa_0'^2
    &=
    \left(u_x+\frac{\lambda}{4}\right)^2+u_y^2+u_z^2,
    \\
    \epsilon_1&=-hu_z,
    \qquad
    \epsilon_2=hu_z^2,
    \qquad
    \epsilon_3=\frac12hu_z(\kappa^2-2u_z^2),
  \end{aligned}
\end{equation}
where $h=\lambda(u_x+\lambda/4)$.
From the exact relation
$(M'_0/M_0)^2
=(m_Q^2+\bm k_\perp^{\prime 2})/(m_Q^2+\bm k_\perp^2)$,
the Jacobian factor expands as
\begin{equation}\label{eq:NR-D-jacobian-cubic}
  \sqrt{\frac{M'_0}{M_0}}
  =
  1+\frac{v^2}{4}\lambda u_x
   +\frac{v^3}{4}\epsilon_1
   +\mathcal O(v^4).
\end{equation}

Combining this Jacobian with the expansion of the Gaussian
exponent gives $w_0^{(1S)}=1$ and
\begin{equation}\label{eq:NR-D-overlap-coefficients}
  \begin{aligned}
    w_1^{(1S)}
    &=-\frac{\epsilon_1}{2},
    \\
    w_2^{(1S)}
    &=-\frac{\epsilon_2}{2}
      +\frac{\epsilon_1^2}{8}
      +\frac{\lambda u_x}{4},
    \\
    w_3^{(1S)}
    &=-\frac{\epsilon_3}{2}
      +\frac{\epsilon_1\epsilon_2}{4}
      -\frac{\epsilon_1^3}{48}
      -\frac{\lambda u_x\epsilon_1}{8}
      +\frac{\epsilon_1}{4}.
  \end{aligned}
\end{equation}
The linear coefficient $w_1^{(1S)}=hu_z/2$ reproduces the
odd recoil correction in Eq.~\eqref{eq:NR-overlap-exp}.
The first three terms of $w_3^{(1S)}$ arise from the
Gaussian exponent, the fourth from the Jacobian--exponent
product, and the last from the cubic Jacobian correction.
All five terms are required in the
$K_{D,-1}w_3^{(1S)}$ contribution to the $\mathcal O(v^2)$
coefficient in Eq.~\eqref{eq:NR-KW-products}.
For the Gaussian averages of Eq.~\eqref{eq:NR-avg-def},
odd moments vanish and even moments factorize, with
$\langle u_a^{2n}\rangle_{\phi_{1S}}=(2n-1)!!/2^n$
for $a=x,y,z$.

The $2S$ and interference overlaps are obtained by multiplying
the Gaussian overlap by the appropriate node factors,
\begin{equation}\label{eq:NR-D-node-coefficients}
  \mathfrak n=2\kappa^2-3,
  \qquad
  \mathfrak n'=2\kappa'^2-3
  =
  \sum_{j=0}^3v^j\mathfrak n'_j+\mathcal O(v^4),
\end{equation}
where $\mathfrak n'_0=2\kappa_0'^2-3$ and
$\mathfrak n'_j=2\epsilon_j$ for $j=1,2,3$.
In particular, $\mathfrak n'_1=-2hu_z$ is the leading
odd recoil correction to the final-state node factor.
The diagonal $2S$ overlap carries
$\mathfrak n\mathfrak n'/6$.
For the interference, the sum of the two ordered radial
products is
\begin{equation}\label{eq:NR-D-cross-overlap}
  \begin{aligned}
    &\widetilde\phi_{1S}^{*}(\vec k\,')
     \widetilde\phi_{2S}(\vec k)
    +\widetilde\phi_{2S}^{*}(\vec k\,')
     \widetilde\phi_{1S}(\vec k)
    \\
    &\qquad=
    \frac{\mathfrak n+\mathfrak n'}{\sqrt6}\,
    \widetilde\phi_{1S}^{*}(\vec k\,')
    \widetilde\phi_{1S}(\vec k).
  \end{aligned}
\end{equation}
The corresponding overlap coefficients follow from
Eq.~\eqref{eq:NR-D-overlap-coefficients},
\begin{equation}\label{eq:NR-D-2S-overlap-coefficients}
  \begin{aligned}
    w_n^{(2S)}
    &=
    \frac{\mathfrak n}{6}
    \sum_{j=0}^n
    \mathfrak n'_j w_{n-j}^{(1S)},
    \\
    w_n^{({\rm cross})}
    &=
    \frac1{\sqrt6}
    \left[
      \mathfrak n\,w_n^{(1S)}
      +\sum_{j=0}^n
       \mathfrak n'_j w_{n-j}^{(1S)}
    \right],
  \end{aligned}
\end{equation}
for $n=0,1,2,3$.

The interference can also be checked by differentiating
with respect to the common Gaussian width.
The normalized HO functions satisfy
\begin{equation}\label{eq:NR-HO-width-identity}
  \widetilde\phi_{2S}(\vec k)
  =
  \frac2{\sqrt6}\,\beta
  \left.
    \frac{\partial\widetilde\phi_{1S}(\vec k)}
         {\partial\beta}
  \right|_{\vec k},
\end{equation}
which follows from
$\beta\partial_\beta\widetilde\phi_{1S}
=(\vec k^{\,2}/\beta^2-\tfrac32)\widetilde\phi_{1S}$.
At fixed $m_Q$, $t$, and $\vec k$, the recoil momentum
$\vec k\,'$ is independent of $\beta$.
Differentiating both factors of the radial product therefore
generates the two ordered interference terms,
\begin{equation}\label{eq:NR-HO-width-product}
  \begin{aligned}
    \beta\partial_\beta
    \left[
      \widetilde\phi_{1S}^{*}(\vec k\,')
      \widetilde\phi_{1S}(\vec k)
    \right]
    &=
    \frac{\sqrt6}{2}
    \left[
      \widetilde\phi_{2S}^{*}(\vec k\,')
      \widetilde\phi_{1S}(\vec k)
    \right.
    \\
    &\qquad\left.
      +\widetilde\phi_{1S}^{*}(\vec k\,')
       \widetilde\phi_{2S}(\vec k)
    \right].
  \end{aligned}
\end{equation}
The fBT kernels and Jacobian are also independent of $\beta$
under these conditions.
Differentiation can therefore be taken through the convolution,
giving Eq.~\eqref{eq:NR-HO-width-cross}.
When this identity is applied to the NR form factors,
both $v=\beta/m_Q$ and $\rho=-t/\beta^2$ must be differentiated.

\subsection{Extraction of $A_q^{\rm NR}(t)$}
\label{Sec:NRQCD-A}

Expanding the common kernel of Eq.~\eqref{eq:A-universal}
at fixed scaled momenta gives
\begin{equation}\label{eq:NR-A-kernel}
  \begin{aligned}
    \mathcal K_A
    &=
    \frac12+\frac{k_z}{2m_Q}+\frac{t}{16m_Q^2}
    +\mathcal O(v^3)
    \\
    &=
    \frac12+v\frac{u_z}{2}-v^2\frac{\rho}{16}
    +\mathcal O(v^3),
  \end{aligned}
\end{equation}
where the second line uses $k_z=\beta u_z$,
$t=-\beta^2\rho$, and $v=\beta/m_Q$.
Comparison with Eq.~\eqref{eq:NR-KW-series} identifies
$K_{A,0}=1/2$, $K_{A,1}=u_z/2$, and $K_{A,2}=-\rho/16$.
Combining these with the overlap coefficients and taking the
Gaussian averages of Eq.~\eqref{eq:NR-avg-def} gives
\begin{equation}\label{eq:NR-A-coefficient-convolution}
  \begin{aligned}
    A_q^{{\rm NR},r}(t)
    &=
    G(\rho)\biggl\{
      \frac12\bigl\langle w_0^{(r)}\bigr\rangle_{\phi_{1S}}
    \\
    &\quad
      +v^2\biggl[
       -\frac{\rho}{16}
       \bigl\langle w_0^{(r)}\bigr\rangle_{\phi_{1S}}
      +\frac12
       \bigl\langle u_z w_1^{(r)}\bigr\rangle_{\phi_{1S}}
       +\frac12
       \bigl\langle w_2^{(r)}\bigr\rangle_{\phi_{1S}}
      \biggr]
    \biggr\}
    +\mathcal O(v^4),
  \end{aligned}
\end{equation}
for $r=1S,2S,{\rm cross}$.
The contributions at orders $v$ and $v^3$ are odd in $u_z$
and vanish after integration.

For the pure $1S$ contribution, the three NLO products
$K_{A,2}w_0^{(1S)}$, $K_{A,1}w_1^{(1S)}$, and
$K_{A,0}w_2^{(1S)}$ have Gaussian averages
$-\rho/16$, $\rho/32$, and $\rho(\rho-8)/512$,
respectively.
Their sum is $\rho(\rho-24)/512$.
In particular, $K_{A,1}w_1^{(1S)}$ contributes because
the product is even in $k_z$, although each factor is odd.
Using the corresponding overlap coefficients for the two
diagonal states yields
\begin{align}
  A_q^{{\rm NR},\phi_{1S}}(t)
  &=
  \frac{G(\rho)}{2}
  \left[
    1+v^2\frac{\rho(\rho-24)}{256}
  \right]
  +\mathcal O(v^4),
  \label{eq:NR-Aq-1S}\\[1ex]
  A_q^{{\rm NR},\phi_{2S}}(t)
  &=
  \frac{G(\rho)}{2}
  \left[
    P_2(\rho)
    +v^2
     \frac{\rho(\rho^3-88\rho^2+2176\rho-13312)}{98304}
  \right]
  +\mathcal O(v^4).
  \label{eq:NR-Aq-2S}
\end{align}

For the interference, the leading node sum is
\be
  \mathfrak n+\mathfrak n'_0
  =
  4\vec u^{\,2}-6+\frac{\rho}{4}.
\ee
Since $\langle\vec u^{\,2}\rangle_{\phi_{1S}}=3/2$,
the Gaussian average removes the first two terms, leaving
$\langle w_0^{({\rm cross})}\rangle_{\phi_{1S}}
=\rho/(4\sqrt6)$.
The interference therefore vanishes at $t=0$, as required
by orthogonality, but survives at finite momentum transfer.
Including the NLO products gives
\begin{equation}\label{eq:NR-Aq-cross}
  A_q^{{\rm NR},{\rm cross}}(t)
  =
  \frac{G(\rho)}{8\sqrt6}
  \left[
    \rho+v^2\frac{\rho^2(\rho-40)}{256}
  \right]
  +\mathcal O(v^4).
\end{equation}
As an independent check, applying the width identity of
Eq.~\eqref{eq:NR-HO-width-cross} to
Eq.~\eqref{eq:NR-Aq-1S} at fixed $m_Q$ and $t$ reproduces
Eq.~\eqref{eq:NR-Aq-cross}.
Here $\beta\partial_\beta G=(\rho/8)G$, while
$v^2\rho=-t/m_Q^2$ is independent of $\beta$.

\subsection{Extraction of $D_q^{\rm NR}(t)$}
\label{Sec:NRQCD-D}

We evaluate Eq.~\eqref{eq:NR-KW-convolution} using
$\mathcal K_D^{(+-)}$ of Eq.~\eqref{eq:Dkernels}
and the overlap coefficients of Sec.~\ref{Sec:NRQCD-W}.
The separate $\mathcal O(v^{-2})$ terms cancel in the
subtraction.
The surviving mass-enhanced term is proportional to
$1-2x=\mathcal O(v)$ and therefore starts at
$\mathcal O(v^{-1})$.
Expanding the full entries of Table~\ref{tab:trace}
at fixed $\vec u$ and $\lambda$ gives the coefficients
of Eq.~\eqref{eq:NR-KW-series},
\begin{equation}\label{eq:NR-D-kernel-coefficients}
  \begin{aligned}
    K_{D,-1}
    &=-\frac{8u_z}{\lambda^2},
    &\qquad
    K_{D,0}
    &=\frac12,
    \\
    K_{D,1}
    &=-u_z\left[
      \frac{4\vec u^{\,2}}{\lambda^2}
      +\frac{2u_x}{\lambda}+\frac34
    \right],
    &\qquad
    K_{D,2}
    &=-u_x^2+\frac{4u_xu_z^2}{\lambda}.
  \end{aligned}
\end{equation}
Here $K_{D,-1}$ and $K_{D,1}$ are odd in $u_z$,
whereas $K_{D,0}$ and $K_{D,2}$ are even.

Combining these coefficients with the overlap expansion
according to Eq.~\eqref{eq:NR-KW-products} gives
\begin{equation}\label{eq:NR-D-coefficient-convolution}
  \begin{aligned}
    D_q^{{\rm NR},r}(t)
    &=
    G(\rho)\Bigl\{
      \bigl\langle
      K_{D,0}w_0^{(r)}
      +K_{D,-1}w_1^{(r)}
     \bigr\rangle_{\phi_{1S}}
    \\
    &\quad
    +v^2\bigl\langle
    K_{D,2}w_0^{(r)}
     +K_{D,1}w_1^{(r)}
     +K_{D,0}w_2^{(r)}
     +K_{D,-1}w_3^{(r)}
      \bigr\rangle_{\phi_{1S}}
    \Bigr\}
    +\mathcal O(v^4),
  \end{aligned}
\end{equation}
for $r=1S,2S,{\rm cross}$, with the Gaussian average
defined in Eq.~\eqref{eq:NR-avg-def}.
The contributions at orders $v^{-1}$, $v$, and $v^3$
are odd in $u_z$ and vanish after integration.
Compared with Eq.~\eqref{eq:NR-A-coefficient-convolution},
the additional products are those involving $K_{D,-1}$.

For the pure $1S$ contribution, the first LO product gives
$\langle K_{D,0}w_0^{(1S)}\rangle_{\phi_{1S}}=1/2$.
The second gives
\begin{equation}\label{eq:NR-D-LO-product}
  \begin{aligned}
    \bigl\langle
      K_{D,-1}w_1^{(1S)}
    \bigr\rangle_{\phi_{1S}}
    &=
    -\frac{4}{\lambda^2}
     \bigl\langle hu_z^2\bigr\rangle_{\phi_{1S}}
    \\
    &=
    -\frac{4}{\lambda}
     \left\langle
       \left(u_x+\frac{\lambda}{4}\right)u_z^2
     \right\rangle_{\phi_{1S}}
    =-\frac12,
  \end{aligned}
\end{equation}
where reflection symmetry gives
$\langle u_xu_z^2\rangle_{\phi_{1S}}=0$ and
$\langle u_z^2\rangle_{\phi_{1S}}=1/2$.
The two LO products therefore cancel after integration,
giving
$\left.D_q^{{\rm NR},\phi_{1S}}(t)\right|_{\rm LO}=0$.
This cancellation requires the even product of the odd
kernel coefficient and the odd recoil correction.

The four NLO products
$K_{D,2}w_0^{(1S)}$, $K_{D,1}w_1^{(1S)}$,
$K_{D,0}w_2^{(1S)}$, and $K_{D,-1}w_3^{(1S)}$
have Gaussian averages
$-1/2$, $-(3\rho+56)/64$, $\rho(\rho-8)/512$,
and $-(\rho^2-32\rho-192)/512$, respectively.
Their sum is $-1$, with the $\rho$-dependent terms canceling
among the last three products.
Thus,
\begin{equation}\label{eq:NR-D-1S}
  D_q^{{\rm NR},\phi_{1S}}(t)
  =
  -v^2G(\rho)+\mathcal O(v^4).
\end{equation}

For the diagonal $2S$ contribution, the node factors change
the overlap weighting and introduce the additional recoil
coefficient $\mathfrak n'_1=-2hu_z$.
Using Eq.~\eqref{eq:NR-D-2S-overlap-coefficients},
the two LO products give
\[
  \begin{aligned}
    \bigl\langle
      K_{D,0}w_0^{(2S)}
    \bigr\rangle_{\phi_{1S}}
    &=
    \frac{P_2(\rho)}{2},
    \\
    \bigl\langle
      K_{D,-1}w_1^{(2S)}
    \bigr\rangle_{\phi_{1S}}
    &=
    -\frac76-\frac{\rho^2}{768}+\frac{\rho}{24}.
  \end{aligned}
\]
In the second average, $-7/6-\rho^2/768$ arises from
the Gaussian exponent correction and $\rho/24$ from
the node recoil.
Adding the two averages cancels the $\rho$-dependent terms
and yields
\begin{equation}\label{eq:NR-D-2S}
  \left.D_q^{{\rm NR},\phi_{2S}}(t)\right|_{\rm LO}
  =
  -\frac23G(\rho).
\end{equation}

The four NLO products
$C_j^{(2S)}
=\langle K_{D,j}w_{2-j}^{(2S)}\rangle_{\phi_{1S}}$,
with $j=2,1,0,-1$, are
\begin{equation}\label{eq:NR-D-2S-NLO-products}
  \begin{aligned}
    C_2^{(2S)}
    &=
    -\frac{\rho^2-64\rho+896}{768},
    \\
    C_1^{(2S)}
    &=
    -\frac{3\rho^3-40\rho^2+2176\rho+78848}{24576},
    \\
    C_0^{(2S)}
    &=
    \frac{\rho(\rho-40)(\rho^2-32\rho+384)}{196608},
    \\
    C_{-1}^{(2S)}
    &=
    -\frac{
      \rho^4-96\rho^3+2752\rho^2-61440\rho-8192
    }{196608}.
  \end{aligned}
\end{equation}
Their sum is $-(\rho^2-44\rho+832)/192$, giving
\begin{equation}\label{eq:NR-D-2S-NLO}
  D_q^{{\rm NR},\phi_{2S}}(t)
  =
  -\frac23G(\rho)
  \left[
    1+\frac{v^2}{128}
      \left(\rho^2-44\rho+832\right)
  \right]
  +\mathcal O(v^4).
\end{equation}
The term $C_{-1}^{(2S)}$ contains the cubic overlap
correction $w_3^{(2S)}$ and is required for the complete
$\mathcal O(v^2)$ result.

For the interference, Eq.~\eqref{eq:NR-D-cross-overlap}
contains the sum of the node factors rather than their
product.
The primed-node recoil correction is therefore not multiplied
by the unprimed node, unlike the diagonal $2S$ case.
The two LO products give
\be
  \begin{aligned}
    \bigl\langle
      K_{D,0}w_0^{({\rm cross})}
    \bigr\rangle_{\phi_{1S}}
    &=
    \frac{\rho}{8\sqrt6},
    \\
    \bigl\langle
      K_{D,-1}w_1^{({\rm cross})}
    \bigr\rangle_{\phi_{1S}}
    &=
    \frac1{\sqrt6}
    \left[
      \left(-\frac{\rho}{8}-2\right)+2
    \right].
  \end{aligned}
\ee
The two terms in the second average arise from the
Gaussian exponent correction and the node recoil,
respectively.
Together with the constant-kernel product, they cancel,
\begin{equation}\label{eq:NR-D-cross}
  \left.D_q^{{\rm NR},{\rm cross}}(t)\right|_{\rm LO}
  =  0.
\end{equation}
This cancellation holds at finite recoil, where
$\langle\mathfrak n\rangle_{\phi_{1S}}
=\langle\mathfrak n'_0\rangle_{\phi_{1S}}=\rho/8$
is nonzero.
It is therefore not a consequence of forward orthogonality alone.

The four NLO products sum to $-(\rho+16)/(4\sqrt6)$,
giving
\begin{equation}\label{eq:NR-D-cross-NLO}
  D_q^{{\rm NR},{\rm cross}}(t)
  =
  -\frac4{\sqrt6}\,v^2G(\rho)
  \left(1+\frac{\rho}{16}\right)
  +\mathcal O(v^4).
\end{equation}
The width identity of Eq.~\eqref{eq:NR-HO-width-cross}
provides an independent check.
Differentiating Eq.~\eqref{eq:NR-D-1S} at fixed $m_Q$
and $t$ gives
\be
  \left.
    \beta\frac{\partial D_q^{{\rm NR},\phi_{1S}}(t)}
         {\partial\beta}
  \right|_{m_Q,t}
  =
  -2v^2G(\rho)
  \left(1+\frac{\rho}{16}\right)
  +\mathcal O(v^4).
\ee
Multiplication by $2/\sqrt6$ reproduces
Eq.~\eqref{eq:NR-D-cross-NLO}.
Holding $\rho$ fixed instead of $t$ would miss the
recoil term proportional to $\rho$.

Substituting Eqs.~\eqref{eq:NR-D-1S},
\eqref{eq:NR-D-2S-NLO}, and~\eqref{eq:NR-D-cross-NLO}
into Eq.~\eqref{eq:NR-Fq-mix-summary} and taking the
forward limit gives Eq.~\eqref{eq:NR-D-mix}.
For mixing coefficients satisfying
$0<c_2/c_1<\sqrt{2/3}$, the momentum-space wave function
\[
  \Phi_{1S}
  =
  \phi_{1S}
  \left[
    c_1+\frac{c_2}{\sqrt6}
    \left(\frac{2\vec k^{\,2}}{\beta^2}-3\right)
  \right]
\]
is nodeless.
Nevertheless, for $c_2$ held fixed as $v\to0$, its leading
forward form factor is
$\left.D_q^{\Phi_{1S}}(0)\right|_{\rm LO}
=-2c_2^2/3\neq0$.
Thus, the leading cancellation for the pure Gaussian
does not follow from nodelessness alone.

As a check of current-component independence, we also
expand the transverse-trace kernel $\mathcal K_D^{(ii)}$.
Unlike the common $(+-)$ and momentum-transfer-contracted
$(-i)$ kernels, it has no $v^{-1}$ term.
Its convolution with the overlap coefficients of
Eqs.~\eqref{eq:NR-D-overlap-coefficients}
and~\eqref{eq:NR-D-2S-overlap-coefficients}
reproduces the LO and $\mathcal O(v^2)$ results for
all three radial contributions.
The $e^i$ contribution of Table~\ref{tab:trace}
is required for this agreement.
Omitting it changes the transverse-trace extraction
but leaves the contracted channels unchanged.

\newpage
%---------------------------------------------------------
% Updated bibliography in the standard colors.
  \bibliographystyle{JHEP}%
  \bibliography{references}%
%
%\bibliographystyle{elsarticle-num}
%\bibliographystyle{apsrev4-2}
%\bibliography{references}
\end{document}